\documentclass[12pt]{article}
\usepackage{scicite}
\usepackage{times}

\newenvironment{sciabstract}{%
\begin{quote} \bf}
{\end{quote}}

\usepackage[T1]{fontenc}
\fontfamily{cmr}\selectfont
\usepackage{titling}
\usepackage{authblk}
\usepackage{array}
\usepackage{titlesec}
\usepackage[letterpaper, total={7.25in, 9.5in}]{geometry}
\usepackage{ amsmath,amssymb}
\usepackage{graphicx,color,transparent,placeins,xr,rotating}
\usepackage{bm,bbm}
\usepackage{hyperref}
\usepackage[labelfont=bf]{caption}
\usepackage[capitalize]{cleveref}
\usepackage[toc,page]{appendix}
\usepackage{xcolor}
\usepackage{soul}
\usepackage{longtable}
\usepackage{enumerate}
\creflabelformat{equation}{#2(#1)#3}

\makeatletter

\newcommand{\beginsupplement}{%
        \setcounter{table}{0}
        \renewcommand{\thetable}{S\arabic{table}}%
        \setcounter{figure}{0}
        \renewcommand{\thefigure}{S\arabic{figure}}%
        }
\DeclareMathOperator{\Tr}{Tr}

\title{Distinguishing Cell Phenotype Using Cell Epigenotype}

\author
{Thomas P. Wytock,$^{1\ast}$ Adilson E. Motter$^{1,2,3}$\\
\vspace{12pt}
\normalsize{$^{1}$Department of Physics and Astronomy, Northwestern University, Evanston, IL 60208, USA}\\
\normalsize{$^{2}$Northwestern Institute on Complex Systems, Evanston, IL 60208, USA}\\
\normalsize{$^{3}$Chicago Region Physical Sciences-Oncology Center, Northwestern University, Evanston, Illinois 60208, USA}\\
\vspace{12pt}
\normalsize{$^\ast$To whom correspondence should be addressed; E-mail:  t-wytock@northwestern.edu.}
}

\date{}

\begin{document}

\baselineskip13pt


\maketitle

\noindent NOTICE: This is the author's version of the work. It is posted here by permission of the AAAS for personal use, not for redistribution. The definitive version was published in Science Advances 6, eaax7798 (2020), doi: 10.1126/sciadv.aax7798.\\

%\noindent {\bf Sentence summary:}
%
%\noindent Gene expression and chromatin conformation reliably distinguish cell types, enabling equation-free control of human cells.\\

\begin{sciabstract}
The relationship between microscopic observations and macroscopic behavior is a
 fundamental open question in biophysical systems. 
Here, we develop a unified approach that---in contrast with existing methods---predicts 
cell type from macromolecular data even when accounting for the scale of human tissue 
diversity and limitations in the available data. We achieve these benefits by applying a 
$k$-nearest-neighbors algorithm after projecting our data onto the eigenvectors of the 
correlation matrix inferred from many observations of gene expression or chromatin conformation.
Our approach identifies variations in epigenotype that impact cell type,
thereby supporting the cell type attractor hypothesis and representing the first step 
toward model-independent control strategies in biological systems.
\end{sciabstract}

%\newpage

\section*{Introduction}

Genetically identical human cells are classified by their distinct behaviors 
into cell types, implying that non-genetic factors---including chromatin 
organization---contribute to their distinctive gene expression patterns. 
Being stably heritable through cell division, both chromatin 
organization and the unique pattern of gene expression are therefore epigenetic\cite{Cortini2016}. 
Observing these epigenetic degrees of 
freedom, or epigenotype, of a wide variety of cells has become increasingly widespread
thanks to technological advances in gene expression microarrays\cite{Shi2006}
and, more recently, genome-wide chromatin conformation capture (Hi-C, which measures 
the genome-wide frequency of physical contact between pairs of loci)\cite{Imakaev2012}
and RNA-sequencing\cite{Ozsolak2010}. 
Collections of data from these experiments are available in
public databases, of which two especially large ones are
 the Gene Expression Omnibus 
(GEO)\cite{Barrett2009} and the Sequence Read Archive (SRA)\cite{leinonen2010sequence}. 
Yet, existing approaches remain too limited in scope to 
distinguish a large number of cell types on the basis of epigenotype, 
hampering the discovery of underlying physical principles that would facilitate 
manipulating cell behavior, cell reprogramming, and developing regenerative therapies. 
On the other hand, statistical physics\cite{Hopfield1982} 
and nonlinear dynamics\cite{Packard1980}, combined with machine learning\cite{Lu2009,Airoldi2016,Wytock2019}, offer a promising 
framework to determine cell type solely from macromolecular data.

Inferring cell type from epigenotype is a challenging problem largely 
because the complexity and scale of the intracellular networks considered here 
(consisting of $10^4-10^6$ genes or gene products) preclude the use of many existing approaches. 
These approaches 
 range from direct simulation\cite{Assaf2013,Lu2014} for networks smaller than 10 genes, 
to Boolean models\cite{Wang2012,Saadatpour2016} for networks up to 100 genes, 
to nonlinear embedding methods\cite{Donner2010,Crutchfield2011} for networks up to 1,000 genes, 
%Crutchfield1989,Toroczkai1993,Voss1998,Feldman2003,Thiel2004,
to inverse Ising models\cite{Lang2014,Dettmer2016} for networks larger than 1,000 genes.
%Cocco2011,Aurell2012,Dettmer2016
Inverse Ising models and other recent network identification approaches\cite{Han2015,Dettmer2016}
must contend with effective interactions between genes, such as those induced by 
the competition for cellular resources\cite{Rondelez2012}, and are sensitive
to missing links in the reconstructed network when making predictions about 
cell behavior. On the other hand, approaches to predict the growth rate of micro-organisms 
from gene expression\cite{Airoldi2016,Wytock2019}, and cell-fate decisions in mice from epigenetic 
markers\cite{Lu2009}, suggest that prediction of cell behavior from whole-cell measurements should
be possible.

Here, motivated by these latter approaches, we present a data-driven approach that benefits 
from machine-learning techniques to infer
cell type based on genome-wide observations of gene expression or 
chromatin conformation without the benefit of a network model. 
The gene-gene or contact-contact correlation matrices provide the structure of the data 
obviating the need for an explicit network model.
We show that our approach
preserves cell type homogeneity
(less variability exists within than between cell types), local 
consistency (states nearby measurements of a cell type likely belong to it), and
data efficiency (state-space regions belonging to cell types may be estimated with few measurements). Applying this approach
to both gene expression and Hi-C datasets, we  
distinguish cell types better than existing methods, 
even when considering a large set of cell types representative of the variety of human normal and cancer tissues.

Assigning gene expression or chromatin conformation states to a phenotype may be regarded as a coding 
problem in information theory, in which we want to choose the set of $L$ binary features 
that most reliably classify the cell types when measurements may be incorrect with probability $t$. 
Here, we use ``feature'' to refer to either a single gene or eigengene, where an 
eigengene is a projection along a single eigenvector of 
correlations between genes. For Hi-C, features are either contacts between pairs of loci or eigenloci,
which are defined analogously to eigengenes.
Non-redundant codes transmit the most information per feature 
transmitted, but are also the most error prone, as the probability of correct transmission scales with $(1-t)^L$. 
In the approach described here, we quantify the cell type homogeneity, local consistency and data efficiency
criteria and use them to identify sets of features that reliably encode cell types. The flexibility of our approach
is apparent from the application to two different methods for characterizing cell state across a diverse collection
of cell types, and its reliability is demonstrated by comparing against other approaches.

\section*{Results}
\subsection*{Dataset description.}
We obtain human gene expression data from GEO\cite{Barrett2009}, all publicly available Hi-C data from
SRA\cite{leinonen2010sequence}, and RNAseq data from the Genome-Tissue Expression database,  referring to these datasets as GeneExp, Hi-C, and GTEx, respectively.
Each dataset $X_{uvi}$ consists of $u \in {1,..,N}$ features, $v \in {1,..,M}$ experiments, and $i \in {1,..K}$ 
cell types where ``experiment'' is used to refer to a single measurement of all features.
 Here ($N$, $M$, $K$) take the values of (17,525, 8,842, 102) for GeneExp, (103,827, 453, 11) for Hi-C, and (20,689, 9,850, 26) for GTEx 
(details available in \hyperref[sec:Methods]{Methods}).  We develop our model on the GeneExp and Hi-C datasets and apply it to the GTEx dataset.
Furthermore, the columns are ordered by cell type such that if $v \in [1+\sum_{k=1}^{j-1} M_k , \sum_{k=1}^{j} M_k ]$ then $i=j$, 
where $j$ is an index over the $K$ cell types and $M_j$ is the number of associated experiments in the dataset.

We investigate the prevalence of correlations in macromolecular data by generating randomly resampled and 
correlated resampled data for each dataset and combine the resampled data with the actual data, as 
described in \hyperref[sec:Methods]{Methods}. In \cref{fig:confusion}, dark blues in the diagonal going from the lower left 
to the upper right represent correct predictions of the state. Note that correlated data often are confused for 
the real data (at a rate of $> 70\%$, much higher than the corresponding rate for uncorrelated data), 
indicated by the blue in the top left and middle left squares, respectively. 
 We further investigate the possibility of using correlations to 
the distinguish cell types in synthetically generated data when they are defined by differences along 
correlation eigenvectors (\cref{fig:method-testing}A, B), 
differences along genes (\cref{fig:method-testing}C, D), 
or a combination of both (\cref{fig:method-testing}E, F).
Concluding that using correlations 
could be advantageous in distinguishing cell types, we choose to construct a redundant encoding based on 
correlations between genes and loci to define cellular state, as indicated in 
\cref{fig:cartoon}A. 

\subsection*{Description of the approach.}
\begin{figure*}[h]
\centering
    \includegraphics[width=\linewidth]{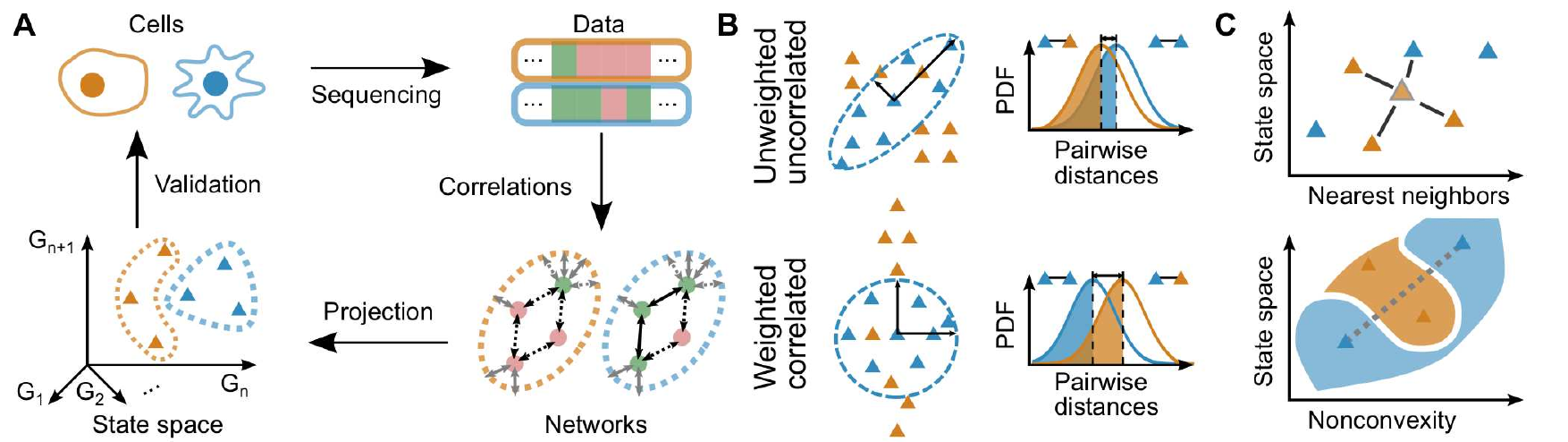}
 \caption{%\baselineskip 10pt 
 \textbf{Schematic illustration of our approach to distinguish cell types.}
 (\textbf{A})The epigenomic measurements of two different cells in blue and orange (top left) yield different epigenotypes (top right) from which an condition-specific effective network (bottom right) is determined from correlations in the data, where solid or dashed lines indicate relationships that are enforced or not enforced but possible, respectively, under the specified conditions. Projection to the state space of correlation eigenvectors approximates the attractors.
  (\textbf{B}) The probability distribution functions of distances between pairs of measurements of the same and different types are compared at selected percentiles (shaded regions) to determine whether pairs of the same type are more similar than pairs of different types. 
  (\textbf{C}) The performance is
 evaluated by using KNN to predict unseen data (top) and by measuring the frequency with which chords cross cell type boundaries (grey dashed line, bottom panel).
  }
  \label{fig:cartoon}
\end{figure*}

We summarize our approach to translate bio-molecular data into cell type in \cref{fig:cartoon}.
Our goal is to select a subset of features that reliably encode cell type.
To this end we formalize the cell type homogeneity, local consistency and data efficiency criteria.

%\subsubsection{Cell type homogeneity}
Cell type homogeneity asserts that there should be less variation within types than between them. 
Thus, we compare the distribution of distances between measurement pairs within a given ``test'' cell type
(labelled with $c$) with the corresponding distributions between the test cell type and all other ``query'' cell types.   
In the test cell type, certain pathways will be active leading to stronger correlations between the constituent 
genes. We propose that after projecting to the correlation eigenvectors, the data should be rescaled in terms 
of the variance along each eigenvector, as we describe next.

Formally, let $\widetilde{X}_{uvi} = (X_{uvi}-\mu_u)/\sigma_u$, 
where $\mu_u$ and $\sigma_u$ are the mean and standard deviation (SD) of each row. 
Using boldface to indicate the suppression of the matrix indices, we decompose the correlations using
\begin{equation}
\bm{\widetilde{X}} = \bm{U}\bm{\Sigma}\bm{V},
\label{eq:SVD}
\end{equation}
where $\bm{U}$ ($\bm{V}$) are the left (right) eigenvectors of the feature correlations and where
 $\bm{X}$ is substituted for $\bm{\widetilde{X}}$ in the case of Hi-C.
Let $c$ be a set of column indices corresponding to a particular cell type (or possibly all cell types), 
and let $\sigma_u^{(c)}$ be the SD of row $u$ calculated using only these indices. Then, we assign 
weights using
\begin{equation}
 \lambda^{(c)}_u = \frac{1/\sigma_u^{(c)}}{\sqrt{\sum_{k=1}^{N}  1/\sigma_k^{(c)\,2}}}. \label{eq:scaling}
 \end{equation}
Noting that $\bm{U} = [G_1, .., G_M]$, where $G_\ell$ are the axes in 
\cref{fig:cartoon}A, we obtain the eigenspace representation of the 
data 
\begin{equation}
\bm{X}' = \bm{U}^\intercal \bm{X} \bm{\lambda}^{(c)}\label{eq:transform}
\end{equation} 
where $\bm{\lambda}^{(c)}$ is a diagonal matrix whose entries are given by \cref{eq:scaling}.
In the weighted versions of our approach, $c$ is the test cell type in one-versus-all classification 
or any cell type in $\{1,.., K\}$ for all-versus-all classification, while in the unweighted version, 
$\bm{\lambda}^{(c)} = \bm{I}$ for all $c$, and in the weighted case 
$\bm{\lambda}^{(c)}$.

To construct the pairwise distance distributions, let $J_1=1$ and $J_i = 1+\sum_{k=1}^{i-1} M_k$ for $i>1$, 
where $M_k$ is the number of experiments associated with the $k$th cell type.
Then the cumulative distribution function of pairwise distances is
\begin{equation}
B^{ij}_S(d) \! =\! \frac{1}{M_{i} M_{j}}\! \sum_{w = J_i}^{J_{i+1}-1} \! \sum_{v = J_j}^{J_{j+1}-1} \! \mathbbm{1}\! \left(|| X’_{\ell w i} \! - \! X’_{\ell v j}||_{\ell \in \{S\}}^2 \! > \! d \right)\!, \! \label{eq:CDF}
\end{equation}
 with the added condition that $w\neq v$ if $i=j$, where $\mathbbm{1}(\cdot)$ is the indicator function, 
 which is $1$ if the argument is true and $0$ otherwise. Let $b^{ij}_S(p)$ be the inverse of  $B^{ij}_S(d)$, where 
 the argument $p \in [0,1]$ is the percentile of the distribution. Then, cell type homogeneity is quantified as
 \begin{equation}
\underset{S}{\min} \sum_{i \neq j} \mathbbm{1}\left( \underset{p}{\max} \left(b^{ii}_S(p) - b^{ij}_S(p)\right) > 0 \right), \label{eq:C1} 
 \end{equation}
 which is illustrated in \cref{fig:cartoon}B. 

%\Cref{eq:C1} is similar to the constraints strongly enforced on a classifier in metric learning\cite{kulis2013metric}. 
In our feature selection procedure, \cref{eq:C1} represents ``soft'' constraint by using it as a 
regularization term in dimension reduction to reflect the possibility that closely (i.e., functionally) related cell types 
have overlapping distance distributions. 

%%\subsubsection{Local consistency}

Local consistency, the idea that the most similar macromolecular profiles should be accurate predictors of cell identity,
undergirds our approach to construct a mapping between genome-wide measurements and cell type.
We proceed by dividing our dataset $X_{uvi}$ into a 
training set $P_{uvi}$ and test set $Q_{lmj}$. 
 For ease of notation, we represent the training set data matrix 
$P_{uvi}$ as ordered pairs of experiments and cell type labels $(\bm{x}_v, i) = P_{uvi}$ with the boldface 
type to indicate that index over the feature labels $u$ is suppressed and $i \in \{1,..,K\}$. Let $D_m$ be 
the set of column indices of the training data matrix corresponding to the $k$-nearest neighbors (KNN) of the test 
experiment and label $(\bm{z}_m, j) = Q_{lmj}$.

Taking $k=9$ ($k=7$ for Hi-C) since we restrict our datasets to have at least 10 measurements 
per cell type, the KNN estimate for the cell type probabilities 
$ \hat{w}^{(c)}_{im} $ of $\bm{z}_m$ is
\begin{equation}
\hat{w}^{(c)}_{im} = \mathrm{KNN}(\bm{z}_m; c) = \frac{ \sum_{n \in D_m}|| \bm{\lambda}^{(c)} (\bm{z}_m-\bm{x}_n)||^2 \delta_{i j}}{\sum_{n \in D_m} || \bm{\lambda}^{(c)} (\bm{z}_m-\bm{x}_n)||^2}, \label{eq:C2}
\end{equation}
 as illustrated in \cref{fig:cartoon}C (top panel). 
The resulting cell type prediction is then $\hat{w}^{(c)}_{m} = \operatorname{arg\,max}_i \hat{w}^{(c)}_{im}$. 

%%\subsubsection{Data efficiency}

For data efficiency, we define a chord between any two measurements of cell type $i$ such 
that $C_{vv’}(s) = (X’_{\ell v’ i} - X’_{\ell v i} ) s + X’_{\ell v}$, where $s \in [0,1]$ and $\ell \in \{S\}$. 
We sample a total of $P = 10,000$ points along each of $Q$ realizations of 
$C_{vv’}(s)$ as follows: let $\kappa=2P//(Q^2-Q)$, where $//$ denotes integer division.  
Then, we randomly select $(\kappa +1) (Q^2-Q)/2 - P$ chords, which are sampled at 
$s=[\frac{1}{\kappa}, .., \frac{\kappa-1}{\kappa}]$, and the remaining chords are sampled at $s = [\frac{1}{\kappa+1}, .., \frac{\kappa}{\kappa+1}]$.
Applying KNN, we obtain $\hat{y}_{a}^{S} = \operatorname{KNN}(C_{vv’}(s); S)$, where $a$ is an index over the $P$ chords. The third criterion is 
\begin{equation}
\underset{S}{\min} \sum_{a=1}^P \mathbbm{1}(\hat{y}_{a}^S \neq i), \label{eq:C3}
\end{equation}
as illustrated in \cref{fig:cartoon}C (bottom panel). %The three criteria are discussed in further detail in the SM,~\cref{global:sec,local:sec,nonconvexity:sec}\cite{Note1}.

The data efficiency criterion (\cref{eq:C3}) estimates the probability that the convex hull defined by the measurements of a cell type also belong to it. 
\Cref{eq:C2,eq:C3} both reflect the fact that cell identity tends to be robust to small perturbations. 
The data decomposition and rescaling, in concert with the KNN method and the three criteria, constitute our approach, which is implemented 
in the source code available at \url{https://github.com/twytock/Distinguishing\_Cell\_Types}.

\subsection*{Comparison with other methods.}
We compare KNN with two other machine-learning techniques, support vector classifiers (SVC) and random forests (RF) to verify that it performs best.
In \cref{tab:comparison}, KNN, SVC and RF are compared with
\begin{table}
\caption{\label{tab:comparison}
Comparison between our machine-learning techniques and existing methods applied to both
 datasets measured by the percentage correct classifications under LOGO cross-validation. }
% Use "S" column identifier to align on decimal point 
\begin{tabular}{l r r }
\hline
 Method &  GeneExp (\%) & Hi-C (\%) \\ 
\hline
KNN &   68.4 &   63.4 \\
SVC &  57.8 &   43.7 \\
RF    &   39.7&   40.0 \\
\hline
HNN &   5.6 &   11.5 \\
PDM &   18.8 &   59.3 \\
\hline
\end{tabular}
\end{table} 
 two other existing methods based 
on Hopfield Neural Networks (HNN)\cite{Lang2014} and spectral clustering 
(PDM)\cite{Braun2011}. In this comparison, we perform all-versus-all classification of cell types. 
In KNN, we calculate $\bm{\lambda}^{(c)}$ in \cref{eq:transform} using all the data as we are performing 
an all-to-all comparison. 
The remaining methods use $\widetilde{X}_{uv}$. Let $v \in \{E_k\},\,k \in {1,..,A}$ be the set of 
columns belonging to a the $k$th GEO Series Accession (GSE), SRA Sequencing Read Project (SRP), or GTEx subject ID. 
To test the accuracy of each method, we take $\{1,..,M\} \backslash  \{E_k\}$ and $\{E_k\}$ as 
training and test sets and compare the predicted and actual cell types for each $\{E_k\}$. We use
 the shorthand ``leave-one-GSE-out'' (LOGO) to refer to this validation strategy, as it reflects the 
 situation of the method being applied to a new experiment about which it has no information,
 as described in the \hyperref[sec:Methods]{Methods}. We note that PDM is an unsupervised method,
 therefore we need to interpret the clusters generated as described in \hyperref[sec:SI]{Supplementary Information}.
 
%\subsection*{Comparison with principal component analysis.}
In addition to the methods in \cref{tab:comparison},
we also compared our approach with that of principal component analysis (PCA), which
 can be used 
 %to find a set of  orthogonal linear combinations in multivariate data that 
to reduce the data dimensionality while maintaining most of the 
variance. As in our approach, the principal components are calculated using
\cref{eq:SVD}, where $\bm{X}$ is the covariance matrix, $\bm{\Sigma}$ are the associated eigenvalues 
and the rows of $\bm{U}$ are the eigenvectors. In PCA, dimension reduction proceeds by finding 
$s$ such that $\operatorname{\arg \min}_{s} \sum_{r=1}^s \Sigma_{rr}/\Tr{\Sigma} > t$. 
Here, the elements of $\bm{\Sigma}$ and associated rows of $\bm{U}$ are ordered by decreasing magnitude, and $t$ is a
threshold representing the fraction of the total variance accounted for by the first $s$ rows. In contrast, our forward 
feature selection procedure, selects sets of features based on an objective function (\hyperref[sec:Methods]{Methods}).

\Cref{fig:pca_comp} shows the 
improvement of our feature selection method on PCA. 
We show 
that the feature sets $S_4$ for GeneExp in \cref{fig:pca_comp}A and $S_3$ for Hi-C in \cref{fig:pca_comp}C perform significantly 
better than their PCA counterparts when 5\%, 10\%, 15\% and 20\% of the data is held out.  
In \cref{fig:pca_comp}B, D,
we show that feature selection converges to the  accuracy it achieves for large feature sets for small 
numbers of features in both datasets. Interestingly, PCA achieves a higher average value for 
the Hi-C dataset for feature sets with more than 15 features, highlighting that the forward 
feature selection procedure can get caught in a local maximum.
Although fewer cell types are represented in the Hi-C dataset, we can still verify that our forward 
selection algorithm outperforms PCA for small numbers of features
in \cref{fig:pca_comp}C, D. 
Because feature selection differs from PCA for both datasets, we suggest 
that the difference from PCA is a generic feature of our overall approach in the context of cell types.

\begin{figure*}[htb]
\vspace{-5mm}
\centering
   \includegraphics[width=\linewidth]{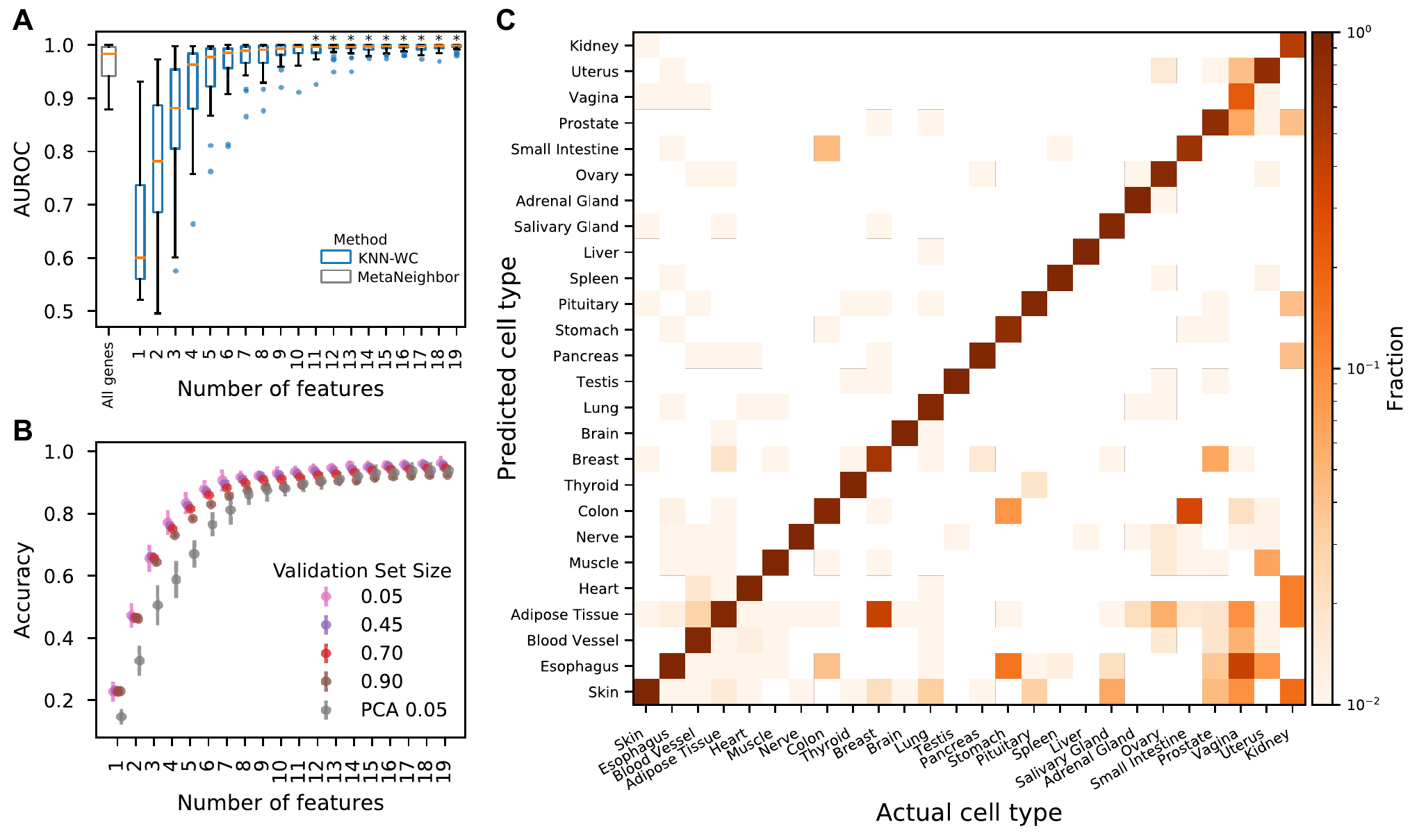}
\caption{ \textbf{Assessment of our method applied to the GTEx dataset and comparison with alternatives.}
(\textbf{A})  AUROC for each cell type presented as a box plot for each number of features. Asterisks indicate significant improvement ($p<0.05$, Kolmogorov-Smirnov test) relative to the MetaNeighbor performance. 
 (\textbf{B}) Accuracy of LOGO validation as a function of the number of features and the size of the test set expressed as a fraction of all experiments. Optimization-selected features perform better than PCA-selected ones, especially for models with few features. 
 (\textbf{C}) LOGO validation accuracy using 9 features, where the cell types are listed in order of the number of experiments.
\label{fig:rnaseq-summary}}
\vspace{-5mm}
\end{figure*}

We also compared the KNN method applied to the GTEx dataset with two methods designed for
single-cell RNA-sequencing: SC3\cite{Kiselev2017} and MetaNeighbor\cite{Crow2018}. The former is
an unsupervised method that achieves 63.0\% accuracy compared with 92.5\% for the KNN method.
The latter is supervised, but its fit criterion is the Area Under the Receiver Operator Characteristic (AUROC), 
which is 0.966 compared with 0.986 for the KNN method (\cref{fig:rnaseq-summary}A). In addition, the KNN method outperforms PCA;
 in particular, a classifier using KNN-selected eigengenes trained on 10\% of the data outperforms a classifier using PCA-selected eigengenes trained on 95\% of the data (\cref{fig:rnaseq-summary}B). The classification rates as a function of predicted and actual cell types 
 %in \cref{fig:rnaseq-summary}C are discussed along with the GeneExp and Hi-C methods.
are presented in \cref{fig:rnaseq-summary}C for the GTEx dataset.

\subsection*{Comparison between versions of the method.}
Because KNN offers superior performance on all datasets, 
we confirm that both the eigenvector representation (\cref{eq:SVD}) and the rescaling 
transformation (\cref{eq:scaling}) are necessary. To achieve this we benchmark 
the previously described WC version against
an unweighted uncorrelated version (UU) and a weighted uncorrelated version (WU). 
The WC version of the KNN technique uses both \cref{eq:SVD,eq:scaling}
against that of a WU version, which uses \cref{eq:scaling} only,
and an UU version which uses neither. 
Note that using an unweighted correlated benchmark is equivalent to UU because they are related by a unitary transformation.
\Cref{tab:version-comparison} demonstrates the superiority of WC to both alternatives in 
distinguishing cell types using both datasets for all three criteria as detailed below.

%\subsubsection{Preservation of homogeneity}
The results reported in the \cref{eq:C1} row of \cref{tab:version-comparison} are broken down by cell type in 
\cref{fig:minsep-breakdown} and \cref{fig:hic-breakdown}A. 
Over 90\% of squares in the plot are gray, indicating that most cell types satisfy the cell type homogeneity criterion. 
In the GeneExp dataset, breast cancers, colon cancers, monocytes, lymphocytes, and leukemias are almost devoid of overlaps
with other cell types for all three versions of the method. The homogeneity as characterized by the WC version, in particular, 
has few overlaps between cancer cell type groups and any other.
Meanwhile, the epithelial cell tissues tend to overlap substantially, reflecting their functional similarity as 
shown in the second row from the bottom and second column from the left in the checkerboard. In addition, neural 
precursor cells (86th row from the bottom) are difficult to distinguish from others, particularly for the 
uncorrelated versions. This lack of distinguishability is consistent with neural precursors' known 
reprogramming capacity as manipulation of a single transcription 
factor is sufficient to induce a pluripotent state\cite{Kim2009}. These findings suggest that our approach is 
preserving aspects of the gene expression space relevant to cell function.

In the Hi-C dataset, the WC version has substantially fewer overlaps in \cref{fig:hic-breakdown}A.
Out of the $110$ comparisons, there are only $7$ overlaps for the WC version. The manner in which cell type heterogeneity 
manifests itself reveals biological similarities.
The UU version shows cell type heterogeneity for types with few observations (top rows) when compared with types with many (left columns).
On the other hand, the WU version fails to distinguish highly variable cell types with a substantial number of measurements,
such as K562.
While the leukemia cell line K562 shows substantial overlap with the other cell lines for all versions, in the WC version two of
those overlaps are with GM12878 and monocyte-derived macrophages (MDM). The latter are a B cell line and a macrophage 
line, respectively. Because the overlaps are all between cell lines derived from white blood cells, the overlap may be due to 
functionally relevant similarities in chromatin structure.

%\subsubsection{Predictive ability}

\begin{figure}[h]
\centering
\includegraphics[width=\linewidth]{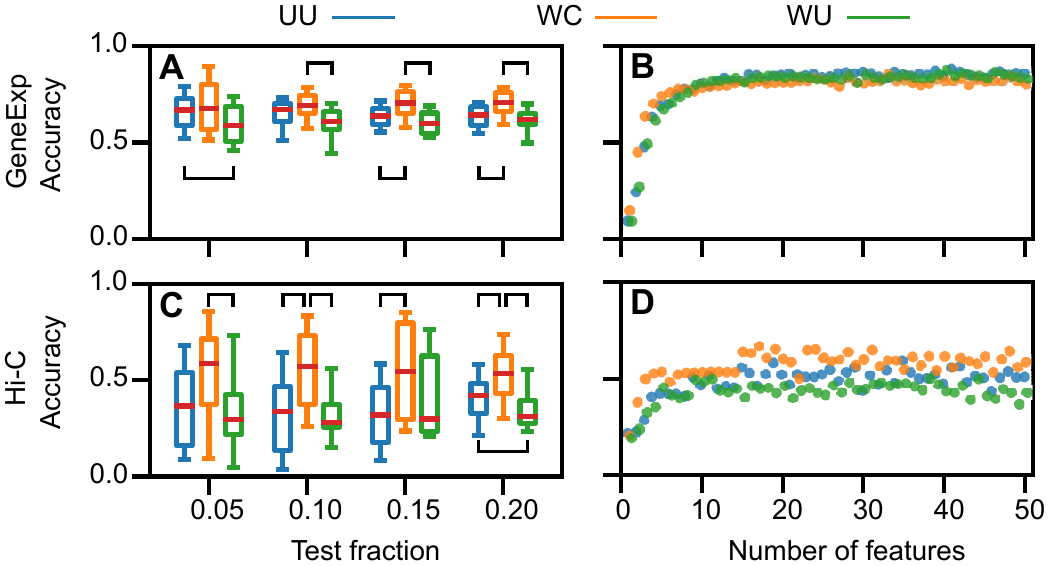}
\caption{\textbf{Comparison of the UU (blue), WC (orange), and WU (green) versions of the KNN technique applied to the GeneExp and Hi-C datasets.}
(\textbf{A}) Boxplots summarizing the distribution of classification accuracy over $n=25$ test sets plotted as a function of the set size indicated as a fraction of all experiments for the GeneExp dataset. Red lines, boxes, and whiskers denote the median, interquartile range and 5th-95th percentile range, respectively.
(\textbf{B}) Mean accuracy plotted as a function of the number of features for the GeneExp dataset.
(\textbf{C}) and (\textbf{D}) Same as (A) and (B), respectively, but for the Hi-C dataset.
 Brackets indicate statistically significant differences between version accuracies as reported in \cref{significance:tab}. }
\label{fig:version_comparison}
\end{figure}

In \cref{fig:version_comparison}, we demonstrate the 
superior predictive ability of the WC version by showing that it maintains its predictive power as 
the size of the test set grows for both datasets. In this case, the test set is constructed 25 times 
by randomly selecting a set of indices $\{g\} \in {1,..,A}$ such that $\sum_{k \in \{g\}} |\{E\}_k| \geq f M$, 
where $f$ is the test fraction, with the constraint that all $K$ cell types are in $\{1,..,M\} \backslash \bigcup_{k \in \{g\}} \{E\}_k$. 
In \cref{fig:version_comparison}A, the WC version performs 
significantly better than the UU or WU versions for test fractions of $0.15$ and $0.20$. 
Similar results are obtained for the Hi-C dataset in 
\cref{fig:version_comparison}C. The WC version is significantly
more accurate than UU for $f=0.05$, more accurate than WU for $0.15$, and more accurate than both for $f=0.10$ and $f=0.20$.

The WC version also achieves higher accuracy when optimizing for a small number of features, as shown in \cref{fig:version_comparison}B, D. 
Letting $\alpha_\ell$ and $\beta_\ell$  be the mean and SD of the accuracy for a KNN model with $\ell$ features and fixed size of the test set, we impose a cutoff for model complexity at
\begin{equation}
\frac{\alpha_{\ell+1}-\alpha_\ell}{\sqrt{\beta_\ell^2 + \beta_{\ell+1}^2}} < 2,\label{eq:termination}
\end{equation}
 resulting in $\ell=4$ for GeneExp, $\ell=3$ for Hi-C, and $\ell=9$ for GTEx. 
  We note that this criterion corresponds to an increase in accuracy being statistically significant at the 95\% confidence level.
 For larger values of $\ell$, the accuracy becomes highly dependent on the construction of the training and test sets, suggesting that the performance of the method is comparable for $\ell > 10$.

\begin{figure*}[htb]
\vspace{-5mm}
\centering
   \includegraphics[width=.7\linewidth]{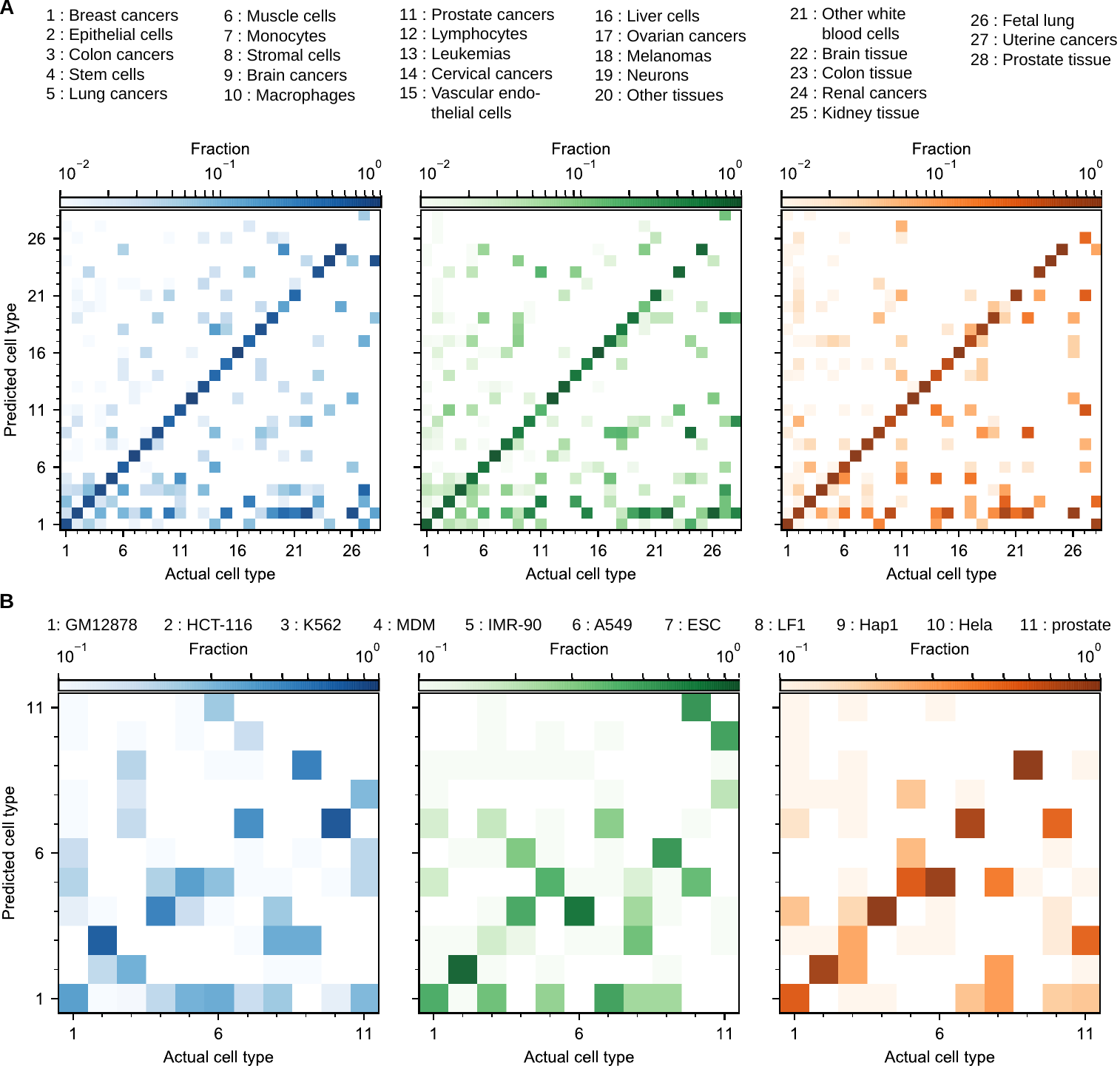}
\caption{\textbf{Comparison of LOGO validation for the three versions of the KNN technique for the GeneExp and Hi-C datasets.}
(\textbf{A}) Validation for the GeneExp dataset using 4 features. The colors indicate the version of the method used to classify the cell types (blue for UU, green for WU, and orange for WC), while the opacity indicates fraction of the total number of experiments belonging to the $x$ axis cell type that are predicted to belong to the $y$ axis cell type.
(\textbf{B}) Same as (A), but for the Hi-C dataset using 3 features.
\label{fig:knn_summary}}
\vspace{-5mm}
\end{figure*}

\Cref{fig:knn_summary} breaks down the KNN results for both datasets when the LOGO validation is used on models of $\ell$ features. 
The cell type groups are ordered left to right and bottom to top by the number of experiments in the dataset in all panels. The presence of 
darker squares along the diagonal in 
the lower left shows that more data make cell types easier to classify.
In \cref{fig:knn_summary}A, experiments are assigned to the correct cell type group with $76.9\%$ accuracy for the WC version, as indicated by the presence of orange color along the diagonal of the right panel. Monocytes (column 7), lymphocytes (column 12), leukemias (column 13), liver tissue (column 16), kidney tissue (column 25), and renal cancer (column 25) are classified without errors using the WC version, reflecting their uniqueness compared with the other cell type groups.
The color in the second row from the bottom in all three panels shows that a variety of cell type groups are often misclassified as epithelial cells, reflecting this cell type group's heterogeneity. In addition, the lack of data for the last three groups in the top right accounts for the method's inability to classify them 
correctly. Under the WC version the brain tissue samples (column 22) tend to be classified as neurons, brain cancers or epithelial cells,
while the remaining missing square along the diagonal corresponds to other tissue sample (column 20), which is a miscellaneous group of
cell types.

For the Hi-C data in \cref{fig:knn_summary}B, 
the classifier maintains accuracy after reducing the entire ensemble of chromatin contacts to 3 dimensions. In the WC version (orange color), seven of eleven cell types
have their largest fraction in a column along the diagonal,
while misclassifications occur between a lung cancer (A549) and two lung cell lines (LF1 and IMR-90, 6th and 8th columns from left, 5th row from bottom).
The misclassification of HeLa cells as embryonic stem cells (ESCs) is interesting, possibly hinting at common replicative potential of both cell lines. Prostate tissue, on the other hand,
has the smallest number of samples in the dataset, making it difficult to classify.

 Results for the WC version of the KNN method applied to GTEx dataset are presented in \cref{fig:rnaseq-summary}C. In this case, 9 features are 
used to classify the cell types reflecting the increased sensitivity of RNAseq as a method compared with gene expression microarrays. Classification
errors are primarily associated with functionally similar tissues (small intestine, stomach, colon, and esophagus) and tissues for which the number of experiments is small.

We break down the results of \cref{tab:version-comparison} and \cref{fig:knn_summary} by cell type in 
\cref{fig:loo-detail}. 
In contrast with pairwise-distance distinguishability, cell types fail to be locally indistinguishable in a less organized 
fashion, reflecting individual measurement variability.
Nevertheless, most misclassifications happen within cell type groups (diagonal of the checkerboard), 
particularly for the WC version of the method, suggesting that when this version misclassifies cell types, 
it often classifies them as a functionally related type. This point is further evidence that the WC version 
is preserving aspects of the gene expression related to cell function. 

The number of experiments of each cell type
%, reported in Table~S2,
impacts how accurately the cell type can be predicted.
\Cref{fig:loo-detail} reveals that the most prevalent 
cell type of each group tends to provoke the majority of the misclassifcations
 (i.e., the top row of each checkerboard row 
and upper diagonal of the diagonal blocks of the checkerboard). This follows 
from the fact that an outlier point in feature space is more likely to be near the most common cell type 
than any other. 
 Taken together, these results support our ansatz that $G_\ell$ and $\bm{\lambda}^{(c)}$ constitute a metric that improves the ability to classify cell types, because changes in gene expression and chromatin conformation must work in concert to effect changes in cell behavior.

%\subsubsection{Attractor reliability}

\Cref{tab:version-comparison} presents the overall fraction chords connecting points of the same cell type that exhibit nonconvexity, 
 with $77.5\%$ of the chords being convex in the GeneExp dataset and $89.5\%$ being convex in the 
 Hi-C dataset for the WC version of our approach. Specifically, 
\cref{fig:hic-breakdown}B  and \cref{fig:ncv-GeneExp}
 break down the fraction of nonconvex chords by cell type. The WC version exhibits greater convexity 
 relative to the UU and WU versions, and with it, more certainty that the interior of the convex hull is part 
 of the cell type. 
% We note that cell types with no secants showing nonconvex behavior are not 
% guaranteed to be convex. However, we are interested in relative nonconvexity of the spaces as opposed 
% to a rigorous proof of convexity. 

 In \cref{fig:ncv-GeneExp}, we see most of the 
 nonconvexity occurring in cell types is structured by cell type group, because the block of pairs with more 
 nonconvex chords than threshold align with the checkerboard boundaries. We observe lung cancers, 
 muscle cells, stromal cells, brain cancers, lymphocytes, melanomas, neurons, fetal lung cells, and uterine 
 cancers all have overlapping chords for the three versions of our approach.

Comparing the smaller Hi-C dataset in 
\cref{fig:hic-breakdown}B with the larger GeneExp 
dataset, we see that the advantage of the WC version becomes more pronounced. 
Only the lung fibroblast cell line LF1, the prostate tissue, and the K562 cell line overlap 
more than two other cell types for the WC version. On the other hand, the WU and UU versions show substantial 
overlaps with the other cell types. Notablly, the IMR-90 cell line does not appear to overlap with LF1 
cell line, despite both of these being developed from lung fibroblasts. Since IMR-90 was 
isolated four decades ago and LF1 was more recently, this lack of similarity in chromatin structure may be 
a side-effect of culturing a cell long term.

Since the overall counts of cell type pairs are not immediately apparent by eye in the preceding figures,
they are enumerated in \cref{fig:overlap-count}. For all three criteria and for both datasets, the WC version has the smallest number of errors.
Because there are fewer cell types in the Hi-C dataset compared with the GeneExp dataset, 
\cref{fig:overlap-count}D has smaller numbers than 
\cref{fig:overlap-count}A.

\section*{Discussion}

The prediction accuracy of $>60\%$ achieved here is greater than what is expected from RNA-protein correlations, 
given that fluctuations in mRNA only account for $< 45\%$ of the variance of the protein abundance\cite{Schwanhausser2011}. 
%We identify protein abundance with cell type based on abundances'  
%  orthologous conservation across taxa\cite{Laurent2010}, 
%  their stoichiometric conservation in multi-protein complexes despite gene copy number variations\cite{Wilhelm2014}, 
% their preservation in response to transcriptional perturbations through post-translational modifications\cite{Liu2016}, 
% and their cell-type-specific profile\cite{Uhlen2015}.
Moreover, given the number of cell types and variables in the GeneExp, Hi-C, and GTEx datasets, 
 it is unclear a priori that machine-learning approaches would work.
 Typically such approaches are developed to classify a small number of distinct items with the number of 
 measurements for each item much larger than the number of variables per measurement. 
 For Hi-C data in particular, the prediction accuracy is surprisingly high 
 since chromatin structure is another step away from protein expression. 
Previous analyses of Hi-C data tend to focus on short-range contacts (i.e., contacts between loci that are $<500$ kb apart), like CTCF-mediated Topologically Associating Domains (TADs), which are often conserved across
 cell types and species~\cite{Tang2015}; they are thus less useful for characterizing cell type compared with alternatives like histone methylation data
 from ChIP-seq\cite{ernst2011mapping,Marco2017}. 
 Nevertheless, our analysis is able to predict cell type from Hi-C data by taking into account the physical interactions between
 distant portions of the genome. This is shown in \cref{fig:hic-distance}A, B, where only contacts between loci respectively greater than or less than a certain distance are included. Understanding the role that chemical interactions play in shaping the long-range physical structure of DNA is an intriguing application case for our method.
Thus, the efficacy of our approach is greater than anticipated from biological and computational considerations.
 
\begin{figure*}
\centering
\vspace{-6mm}
\includegraphics[width=.8\linewidth]{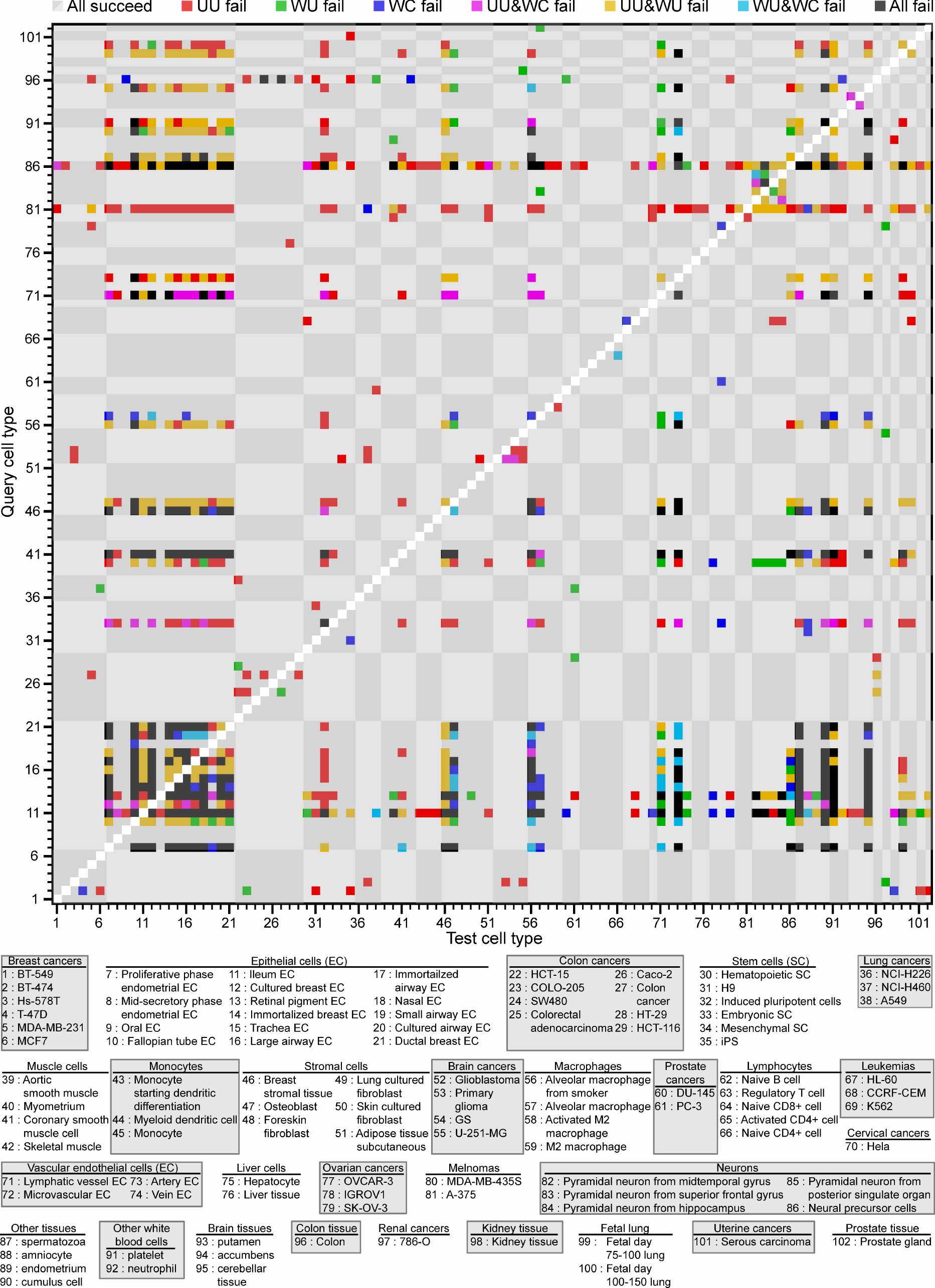}
\caption{ \textbf{Distinguishing cell types by the cell type homogeneity criterion for the GeneExp dataset.}
\Cref{eq:C1} quantifies the cell type homogeneity according to the unweighted 
UU, WU, and WC versions of measuring 
distance. The grey and white checkered background corresponds to the cell type groupings, 
%enumerated in Table~S2, 
and tick labels indicate the cell type associated with
each row and column based on the key below the figure.
The color-coding defined in the legend above the figure marks the cases in which one or more of the 
versions failed for each query (row) and test (column) cell type. Grey indicates that the 
identification was successful for all three versions ($91.4\%$ of all cases). Self-comparisons (white diagonal) were not evaluated.} % with the remaining cases quantified in \cref{fig:overlap-count}B
\label{fig:minsep-breakdown}
\end{figure*}

The success of the approach in reducing dimension to $3$-$9$ features while maintaining predictive accuracy
has several biological implications.
%Our approach achieves this accuracy while simultaneously reducing dimension , 
First, this success stands in contrast to the previous lack of success in establishing biomarkers 
that reliably identify cancer subtypes\cite{Haury2011}. 
Second, the selected features differ from those selected by PCA in that they 
contain subdominant eigenvalues of the correlation matrix, reflecting the multi-scale nature of cell type 
(\cref{fig:rnaseq-summary}B and \cref{fig:pca_comp})
 arising from the hierarchical character of differentiation. This is 
particularly pronounced in 
\cref{fig:rnaseq-summary,fig:minsep-breakdown,fig:loo-detail,fig:ncv-GeneExp}, 
where misclassifications tend to cluster between similar cell types, particularly for the WC approach.
These trends suggest that closely related cell type are being distinguished by subtle changes 
(features associated with smaller eigenvalues) and distant cell types are being distinguished 
by broader changes (features associated with larger eigenvalues).
Third, the inclusion of smaller eigenvalues, which contribute to noise 
 sensitivity in so-called ``sloppy'' models\cite{Transtrum2011}, %Brown2003
 reinforces the 
 necessity of using larger datasets and highlights the shortcomings of PCA in terms of distinguishing 
 cell type.
Fourth, successful dimensional reduction suggests that the selected features constrain variability in the 
unobserved cellular degrees of freedom\cite{yang2012network,Wytock2019}, which is consistent with
%the claim that  
previous equation-free nonlinear embedding methods that distinguish 
network behaviors\cite{Donner2010,Crutchfield2011}.
%Crutchfield1989,Toroczkai1993,Voss1998,Feldman2003,Thiel2004,
The compression of genome-wide data to 3-9 features is also consistent with the observed small 
scaling exponent between the number of cell types and the number of genes\cite{Bell1997,Bonner2004}, 
and supports the hypothesis that cell types are attractors of the underlying intracellular network dynamics\cite{Huang2005}.
%Takahashi2007,Enver2009 
The empirically determined convex hull approximates the basin of attraction of the cell type attractor. 

The successful application of our method to (protein-coding) RNAseq data across a diverse set of tissue types reflects both its accuracy and flexibility. 
The increase in predictive power from 76.9\% in the GeneExp dataset to 92.5\% in the GTEx dataset suggests that most misclassifications are attributable to the less sensitive nature of microarray experiments compared with RNAseq. Second, the favorable comparisons between our method and others applied to 
the GTEx dataset strengthen the conclusion that our method can predict cell type better than existing ones. Furthermore, application of our method to three datasets suggests that it could also be applied to non-coding RNAs to understand their functional role of in shaping cell types\cite{Sarropoulos2019}. 
It is possible to use the annotations to attribute functional information by masking the information for specific sets of genes and observing the change in predictive accuracy (\hyperref[sec:SI]{Supplementary Information}). The success of our method also demonstrates its expected ability to interpret  phenotype in forthcoming experiments in the context of large databases of existing cell type patterns.
%Furthermore, our method could also be applied to understand the role of non-coding RNAs in shaping cell types\cite{Sarropoulos2019}.
%Our approach is drawn from ideas in statistical physics and complex networks.
%Holding a subset of the $G_\ell$ fixed derives from the maximum entropy framework\cite{Jaynes1957}. 
%Assuming the cell type attractor hypothesis to be valid, 

Here, we showed that the correlation decomposition in \cref{eq:SVD} and rescaling factors
in \cref{eq:scaling} increase the fraction of points in the convex hull identified with the cell type in 
\cref{fig:hic-breakdown}B and \cref{fig:ncv-GeneExp}.
These transformations are motivated by the usage of a nonlinear transformation to improve the 
convexity of predictions of network properties from data\cite{Horvat2015}.
In other words, these transformations enhance the resolution of the basin.
Thus, our approach 
offers a solution to the challenging problem of 
estimating basins of attraction for high-dimensional systems from data
and provides evidence for the notion that cell types are identifiable from 
genome-wide expression or chromatin conformation in spite of the high 
dimension of these measurements. %(SM,~\cref{nonconvexity:sec}\cite{Note1}).

Two additional opportunities derive from our approach. First is the development of a 
 semi-supervised version that could identify previously unrecognized cell behaviors, 
 using ideas from refs.\cite{Braun2011,Kiselev2017}.
% , using ideas from refs.~\cite{Schreiber1997,Kraskov2004,Lam2006}. 
 Second is the  
 application of manifold discovery techniques like 
 $t$-distributed stochastic neighbor embedding
 ($t$-SNE)\cite{Maaten2008} 
 %or uniform manifold approximation and projection (UMAP)\cite{McInnes2018} 
 to further refine the selected features and enhance 
 data visualization. 
 %Third is the further development of geometric control algorithms\cite{Toroczkai1994} and equation-free methods to assess basin stability\cite{Menck2013,Mitra2017}.

Our approach advances the field of network medicine, which seeks to integrate large bioinformatic datasets
to direct research into disease treatment\cite{Sonawane2019}.
The global scope of our approach, in tandem with the resulting evidence for the cell type 
attractor hypothesis is the first step in developing
model-independent control strategies in cellular networks. 
Such strategies consist of identifying cell type attractors, curating the macromolecular responses of the cell to 
perturbations, and finding combinations of these responses that together steer the cell from one attractor to another.
Thus, in addition to distinguishing cell types based solely on genome-wide measurements, our approach
could orient the development of rational strategies for 
cell reprogramming, the identification of therapeutic interventions, and other applications 
involving a combinatorially large number of options.

\section*{Methods}
\label{sec:Methods}

\subsection*{Data preprocessing.}
All of the data used in this study is publicly available on the GEO and SRA databases maintained by the NIH 
(for a list of accession numbers, see source code%\cite{Note2}
). For the gene expression data, we chose to look for experiments conducted on the 
Affymetrix HG-U133+2 platform (GPL570 GEO accession), because its use was widespread and the probes 
could be mapped to the hg19 build of the human genome. We applied five different filters for experiments 
on this platform:

\begin{itemize}
\item[(I)] We searched for experiments in which genes were perturbed to gain insight into how the cellular network processes information and thus to infer how the genes are correlated with one another under different conditions.
\item[(II)] We chose to gather gene expression assays using the NCI-60 cell lines as a proxy of human cancers
because these cell types are commonly available and used to screen drugs and other compounds for their 
effectiveness in treating cancer. 
\item[(III)] We obtained gene expression data from the Cancer Cell Line Encyclopedia to sample a wider variety of cancer cell lines.
\item[(IV)] We also included ``normal'' cell types and intermediate cell states obtained by searching for reprogramming experiments.
\item[(V)] We retrieved data from a study that attempted to identify transcription factors that control cell identity to broaden the spectrum of cell types included. 
\end{itemize}
Data from source (I) was used only for the purpose of constructing the correlation matrix, 
while only unperturbed cells in sources (II--V) were used to train and validate the model.
Taken together,
 the combined
dataset, collectively referred to as ``GeneExp,'' comprises 102 distinct cell types 
with $ > 10$ observations. 
We downloaded the raw data from GEO and preprocessed it with a custom CDF file based on the hg19 build 
of the human genome to select the probes that correspond to genes\cite{Dai2005}. After preprocessing the 
gene expression using Robust Median Averaging\cite{Irizarry2003}, we ``batch corrected'' the data, which 
attempts to filter out systematic experimental effects\cite{Johnson2007}. 
%The accession numbers are listed in Table~S2.

The chromatin conformation data, referred to as ``Hi-C,'' was also obtained from 
GEO/SRA by searching for ``Hi-C'' or ``HiC'' while filtering the organism to \emph{Homo sapiens} (Access date: 
 September 25, 2018). 
The files were iteratively corrected as described previously\cite{Imakaev2012}, using the tools available at 
\url{https://github.com/mirnylab/}. 
Chromosomal contacts were binned at 100 kb resolution so that experiments with lower resolution 
sequencing coverage could be included. 

The RNAseq data, referred to as ``GTEx,'' was obtained from the GTEx Portal website: \url{https://www.gtexportal.org/home/}. We used the version 8 gene count data and associated annotations. For RNAseq data derived from lysate or cell pellet, the data were normalized for the total number of reads in each experiment, filtered to include only genes form high-quality experiments (SMATSSCR$<2$) that were expressed in $>1\%$ of all experiments at $>10$ times the minimum expression level. The remaining data were log-transformed and batch corrected based on the SMGEBTCH identifier. We also filtered out any cell types with fewer than 8 experiments. The preprocessing according to the described criteria resulted in 9,850 samples with 20,689 gene identifiers representing 26 cell types (SMTS identifier) from 980 subjects.

The data processing results in gene 
expression levels which have SDs approximately independent of their mean, making decomposition 
of $\bm{\widetilde{X}}$ advantageous.
However, the distribution of Hi-C counts have a long tail because nearby loci come 
into contact exponentially more frequently than distant loci, so decomposing $\bm{X}$ rather than $\bm{\widetilde{X}}$ 
 in \cref{eq:SVD} is more appropriate as the SD of interlocus contacts are dependent on the mean. 

\subsection*{Testing correlation predominance.} 
%To motivate our use of correlations, we show that correlation eigenvectors preserve the structure of biological data in Fig.~S1.
To test whether correlations imparted noticeable structure in each dataset, a classifier was trained 
on a dataset consisting of the actual data, uncorrelated random data, and correlated random data.
First, uncorrelated and correlated data were generated by randomly permuting actual measurements of each 
feature before and after projecting onto the eigenvectors, respectively, resulting in a dataset consisting of $M$ 
instances of each experimental observations, uncorrelated simulated profiles, and correlated simulated profiles.
From this set of $3M$ instances, a training set of size $2M$ was drawn, 
comprising one third of each real data, uncorrelated data, and correlated data, with each instance labeled based 
on how it was generated. A KNN classifier trained on this data predicted the generation method for each profile 
in the test data. For the GeneExp dataset, classification was performed using correlation eigenvectors with an associated eigenvalue $\lambda>1$,
a total of $1,063$, to reduce the impact of noise.

 We also explored our method's performance on synthetic data as a function of the signal-to-noise ratio (SNR). 
Simulated data for 100 ``genes'' from 2 ``cell types'' were generated
using a fixed correlation structure derived from randomly resampling the correlations from the GeneExp dataset.
Cell types were distinguished by introducing differences into randomly selected genes/eigengenes at SNRs ranging from 0.05 to 20.
All models were trained on a small set of fixed size and evaluated on a validation set of 10,000 randomly generated profiles. Both training and
test sets had equal numbers of cell types. 
This analysis was performed for three cases:
\begin{enumerate}[I]
\item \textit{Eigengene-defined cell types.} The data for 1-4 (randomly selected) eigengenes was perturbed by adding the SNR$/2$ to one cell type and subtracting it from the other.
\item \textit{Gene-defined cell types.} The data for 1-4 genes was perturbed by adding the SNR$/2$ to one cell type and subtracting it from the other.
\item \textit{Correlation-defined cell types with confounding genes.} The data for one eigengene is perturbed at an SNR of $5$ and the data for one gene is perturbed by cell type in the training set only, at the prescribed SNR.
\end{enumerate}
 The the size of the training set per cell type was 2, 3, or 5 in cases (I) and (II) and 3, 4, or 5 in case (III), which are the smallest
numbers required to distinguish cell types in each case. 
In cases (I) and (II), the SNR is simply the nominal SNR used to generated the data. 
In case (III), the SNR$=5/\max{|\mu'_i - \mu_i|}$, where $\mu_i$ is the mean difference between the cell types along eigengene $i$ before the gene perturbation is added and $\mu'_i$ is this quantity after it is added, as reported in \cref{fig:method-testing}E, F. This accounts for the instances in which an eigengene with a small associated eigenvalue is dominated by the gene perturbation.

\subsection*{KNN cross-validation.}
In our cross-validation analysis,
we apply \cref{eq:C2} to each experiment of the test set and compare the resulting predictions $\hat{w}^{(c)}_{jm}$ 
with the known measurements $w_{jm}$. We use a one-versus-all classification scheme in which the test cell 
type had one label and the remaining cell types had another because $(\hat{w}^{(c)}_{jm} - \delta_{cj})^2$ is 
the same regardless of the number of remaining cell types.

We adopted different standards for \cref{eq:C2} and strategies for choosing the test set and training set depending 
on whether we were performing cross-validation or testing our approach's performance on unseen data. In the case 
of cross-validation, we adopt the one-versus-all standard, in which measurements of a cell type were assigned to 
the test class and all remaining measurements assigned to the query class.

\begin{itemize}
\item Cross-validation proceeded by dividing the dataset into three pieces, called ``folds,'' constrained to be 
equal both in size and in the ratio of test to query cell types. We cross-validate three times, once with each 
fold as the test set. We then calculated $\sum_{c=1}^{K} (\hat{w}^{(c)}_{jm} - \delta_{cj})^2$ to evaluate the 
overall accuracy across all single cell type frames of reference. This cross-validation scheme is employed in 
the feature selection method.% described in \cref{feature-selection:sec}.,

\item For the purposes of Figs.~\ref{fig:hic-breakdown}a, \ref{fig:loo-detail}, and \ref{fig:overlap-count}A, D,
we employ the LOGO strategy under the standard of all-versus-all classification. In this standard, each cell 
type is assigned to its own class. In the figures, we color the block white if $1/M_{j} \sum_{m=J_j}^{J_{j+1}} \hat{w}^{(c)}_{jm},\ c\neq j < R$, 
with threshold $R = 0.05$ for GeneExp and $R = 0.1$ for Hi-C. Note that some GSE/SRPs contain 
all of the observations of a given cell type. In such cases, the KNN method automatically fails for that 
cell type under that particular test set. Therefore, our success could be even higher if we had restricted 
to cases where all cell types is available in the training set. 

\item The largest GSE/SRP/Subject ID was $<5\%$ of the data, so we extended the LOGO 
procedure to construct larger test sets as reported in 
\cref{fig:version_comparison}A, C and 
\cref{fig:pca_comp}A, C. For these figures, we 
randomly selected GSEs/SRPs until the test set was at least as large as the desired fraction 
$f \in \{0.05, 0.10, 0.15, 0.20\}$ (or up to 0.90 in the case of the GTEx dataset) with the restriction that all cell types must be represented in the training data.
\end{itemize}

\subsection*{Feature selection.} 
Our framework is a hybrid of metric learning and supervised learning 
techniques. %kulis2013metric,
Thus, our objective function consists of a loss term based on the accuracy of the classifier described 
in \cref{eq:C2} and a regularization term based on \cref{eq:C1}. To make explicit the impact of the set of features 
$S$, we rewrite \cref{eq:C2} as
\begin{equation}
\hat{w}^{(c,S)}_{im} = \mathrm{KNN}(\bm{z}_m; c, S) = 
 \frac{ \sum_{n \in D_m}  || \bm{\lambda}^{(c)} (z_{\ell m}-x_{\ell n})||_{\ell \in \{S\}}^2 \delta_{i j}}{\sum_{n \in D_m} || \bm{\lambda}^{(c)} (z_{\ell m}-x_{\ell n})||_{\ell \in \{S\}}^2} \label{eq:KNN}
\end{equation}
and note that $S$ is already explicit in \cref{eq:C1}.
Letting $r_{S}^{ij} = \mathbbm{1}\left(b_{S}^{ii}(p)>b_{S}^{ij}(p)\right)$, our objective function is
\begin{equation}
F(S)=\sum_{c \in \{T\}} \sum_{m \in \{M\}} (\hat{w}^{(c,S)}_{cm} - \delta_{cj})^2 + \gamma \sum_{i,j}^{i\neq j}\sum_p^{\{0,0.5,1\}} r_{S}^{ij} (p), \label{eq:objective}
\end{equation}
where $\gamma = 0.5$ is a scalar regularization parameter that controls the strength of \cref{eq:C1}, 
giving it approximately half the importance of \cref{eq:C2}. Values of $\gamma \gg 1$ will select features 
solely based on satisfaction of the \cref{eq:C1}, while $\gamma \ll 1$ will ignore this requirement in favor 
of \cref{eq:C2}.

With the objective function defined, we describe the forward feature selection algorithm. Recall that $N$ 
is the number of features of the dataset. We first define $\{U_1\} = \{\{i\},\, i \in \{1,..,N\}\}$. Our scheme 
for dimension reduction proceeds by finding $S_1 = \operatorname{arg\,min}_{\{S \in U_1\}} F(S)$, then 
constructing $\{U_2\} = \{ \{S_1,\, i\}\, \, i \in \{1,..,N\}\backslash S_1 \}$. Continuing iteratively, sets of 
features of arbitrary length $S_\ell$ may be constructed. We continue until $\ell=50$, which is long after 
the addition of features has stopped improving the classification accuracy in the LOGO tests (\cref{fig:version_comparison}).

\section*{Acknowledgments}

 This work was supported by CR-PSOC Grant No.\ 1U54CA193419. 
TPW also acknowledges support from NSF-GRFP fund No.\ DGE-0824162 as well as 
NIH/NIGMS 5T32GM008382-23.

\section*{Author contributions}
TPW and AEM designed the research, and wrote and edited the manuscript.
TPW wrote the software, curated the dataset, and analyzed the results.

\section*{Competing interests}
The authors declare that they have no competing interests.

\section*{Data availability}
All analyses in this paper were conducted on data that is publicly available through GEO, SRA, or the GTEx consortium.
Lists of accession numbers may be found either on the GitHub repository (\url{https://github.com/twytock/Distinguishing\_Cell\_Types}) or through the ``Table S2'' XLSX file accompanying the published version of this paper
(\url{https://advances.sciencemag.org/content/suppl/2020/03/16/6.12.eaax7798.DC1}).

\newpage

\section*{List of Figures}

%FIG1-Cartoon

%FIG2-GTEx results
\newpage

%FIG3-pairwise-distance distinguishability

\newpage

%FIG4-LOGO comparison and dimension reduction
\FloatBarrier

\newpage

%FIG5-LOGO cross validations for the three versions for both GeneExp and Hi-C

\newpage
\clearpage
\appendix
\beginsupplement 
\section*{Supplementary Information}
\label{sec:SI}
\subsection*{Method testing for synthetic data.}
\begin{figure*}[htb]
\vspace{-.75cm}
\centering
\includegraphics[width=.9\linewidth]{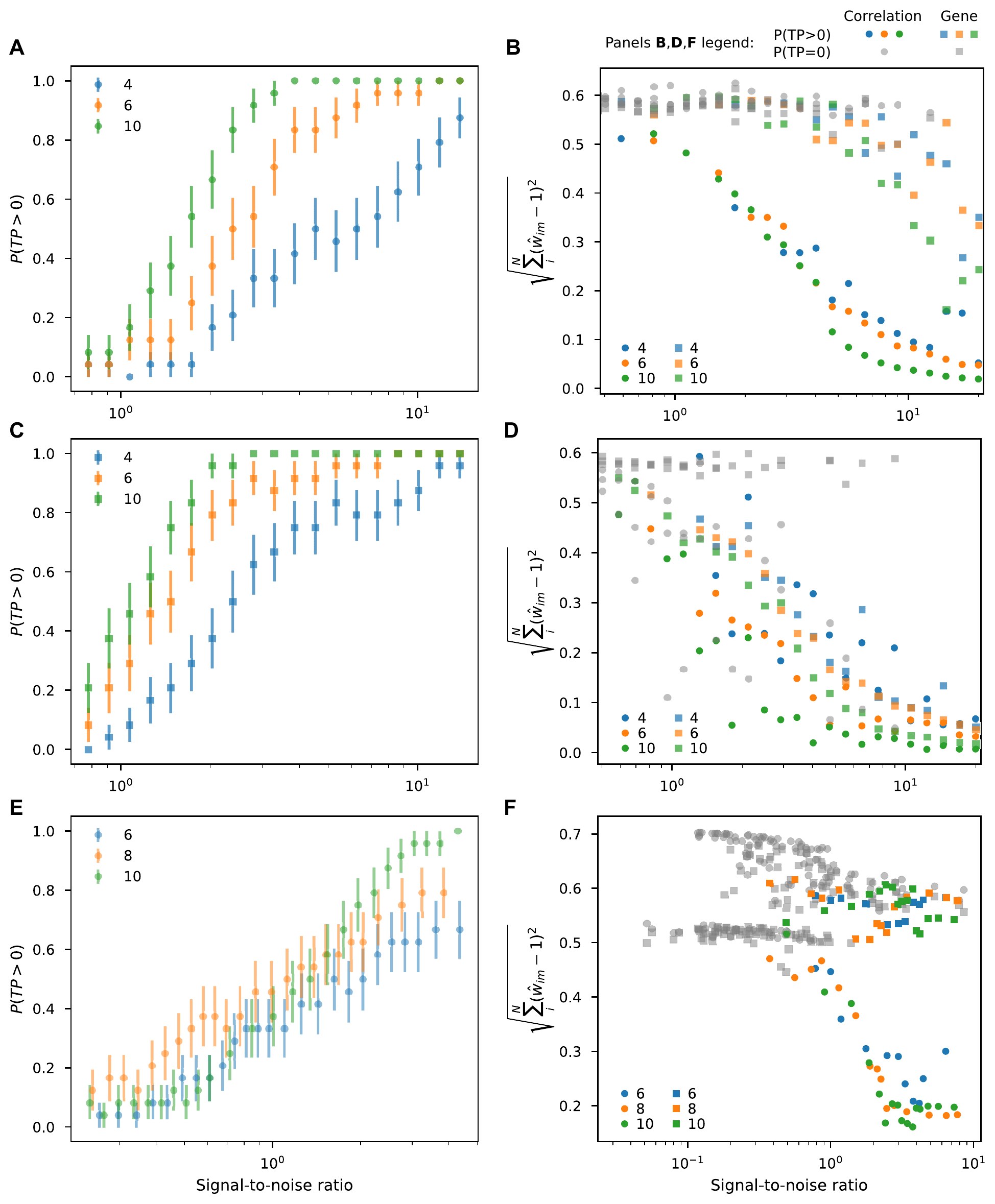}
 \captionof{figure}{
\textbf{Method testing results as a function of the SNR under three scenarios (rows) for two criteria (columns).} 
(\textbf{A}) Probability of identifying a differentially expressed eigengene as a function of the SNR and the number of experiments (color-coded), with error bars denoting the standard error of the mean, for eigengene-based cell types.
(\textbf{B}) Root mean square deviation between the KNN-inferred probability $\hat{w}_{im}$ and the actual cell type as a function of SNR. 
Instances in which differentially expressed genes or eigengenes are not identified are colored in gray.
(\textbf{C} and \textbf{D}) Results for gene-based cell types. Axes, colors, and symbols are as defined in (A) and (B), respectively.
(\textbf{E} and \textbf{F}) Results for eigengene-based cell types with one confounding differentially expressed gene between the cell types in the training set. 
Axes, colors, and symbols are as defined in (A) and (B), respectively.
}
\label{fig:method-testing}
\end{figure*}

We generated synthetic data as described in \hyperref[sec:Methods]{Methods} under three scenarios to test the 
efficacy of the KNN eigengene method versus the gene-based method. The first scenario, reflected in 
\cref{fig:method-testing}A, B, 
shows that the correlation-based method does indeed perform better than the gene-based method. In (A), the 
rate at which the correlation method correctly identifies at  least one of the cell-type defining eigengenes climbs 
steeply between as the SNR varies from 1 to 10. In (B), the correlation method, but not the gene method, 
accurately identifies cell types over this SNR range, as measured by the root mean squared error between 
the KNN-predicted probability of cell type membership ($\hat{w}_{im}$) and the actual cell type measurement. If the probability 
of belonging to a cell type were uniformly distributed between 0 and 1, this measurement would converge to 
$\sqrt{2}$, while perfect prediction corresponds to 0. The deviation from $\sqrt{2}$ for small SNR is attributable to 
the feature optimization step, as it selects the best feature out of a set of 100. We note that 
incorrect identification of one of the eigengenes tends to penalize model accuracy. This reflects eigengenes with 
small eigenvalues being selected to define the difference between cell types, resulting in the effective SNR 
being much smaller than the nominal SNR.
\Cref{fig:method-testing}C, D shows that the correlation method still manages to perform well, 
even when the genes define
the cell type in lieu of the eigengenes. While the probability of identifying the correct gene increases faster than
 the probability of identifying the correct eigengene, the gene-based method fails more drastically when it cannot 
 identify the correct gene. The superior performance of the correlation method in this case is 
 explained by the fact that the difference in a given gene is distributed across all of the correlation eigenvectors 
 so that the method is not sensitive to whether the correct gene is deduced or not. Finally, we consider the case in which 
 cell type differences are defined by a correlation difference, but a single gene is spuriously differentially 
 expressed in the training set, but not in the test set (\cref{fig:method-testing}E, F). When the 
 correlation eigenvector can be identified, which happens almost as often as in (A), the correlation method 
 uniformly outperforms the gene-based method. Thus, the correlation-based method is robust to single-gene errors 
 and can work even in cases where the cell types are defined by genes. We note that our method performs 
 well because there is an underlying correlation structure to the data in all cases, which is a well grounded 
 assumption for biological systems. In contexts where the underlying variables are \textit{uncorrelated}, we would expect 
 the performance of the correlation-based method to deteriorate relative to single-feature methods.

\subsection*{Assessing unsupervised methods.}
Since PDM\cite{Braun2011} and SC3\cite{Kiselev2017} are unsupervised methods, the number of clusters $C$ 
that it produces is not constrained to be equal to the number of cell types $K$. If $C \ll K$, 
(as is the case in which PDM is applied to the GeneExp dataset), then PDM will necessarily have 
limited accuracy, but we can assign a cell type to each cluster by determining the cell type of 
the largest fraction of measurements belonging to that cluster. However, in the case that $C = M$, 
where $M$ is the number of experiments, assignment of each cluster to the cell type of that 
experiment would rate as ``perfect’’ prediction. Such a partition is uninformative. Thus, simply 
assigning each cluster to the largest fraction cell type will overstate the method’s accuracy when the 
clustering method subdivides the experiments to many groups.

Therefore, we calculate the accuracy using the following thought experiment. Suppose that 
we sample one experiment from each cluster, chosen at random, and determine the cell type. 
Then $p_i^{(k)} = m_i^{(k)}/m^{(k)}$ is the probability of sampling cell type $i$ in cluster $k$, 
where $m^{(k)}$ is the total number of measurements in cluster $k$, of which $m_i^{(k)}$ 
belong to cell type $i$. The average number of experiments correctly predicted in the cluster is 
\begin{equation}
 n^{(k)} = \sum_{i \in \{k\}} p_i^{(k)} m_i^{(k)}, \label{eq:num_predict_clust}
\end{equation}
where \{k\} is the set of cell types in cluster $k$. The total fraction predicted is then
\begin{equation}
h = \frac{1}{M} \sum_{k \in C} n^{(k)}. \label{eq:frac_predict}
\end{equation}
Suppose further that each cell type is only able to be assigned once, then \cref{eq:num_predict_clust} 
remains the same, but \cref{eq:frac_predict} must be modified so that no cell type is assigned to 
two different clusters. We look for the best assignment of cell types to clusters by randomly 
assigning cell types to each cluster, and continue until either all clusters have a cell type or 
all cell types have been assigned. We repeat this 1,000 times and take the maximum value 
found as the method accuracy.

\subsection*{On the role of long-range contacts in Hi-C data.}
\begin{figure*}[htb]
\centering
\includegraphics[width=.67\linewidth]{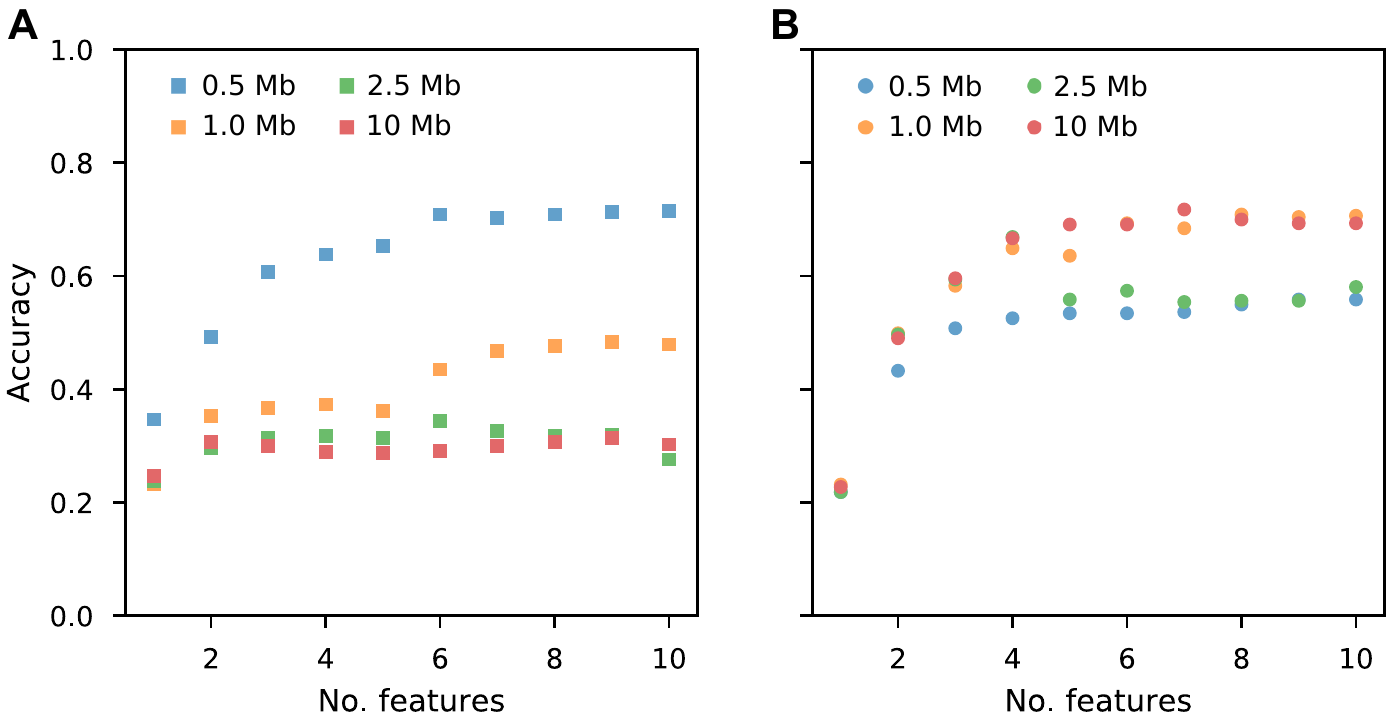}
 \captionof{figure}{
\textbf{Accuracy as a function of the genomic distance between loci and the number of features for the Hi-C dataset.} 
(\textbf{A}) Accuracy when classifying on loci pairs separated by more than the distance in megabases (Mb) indicated by the legend. 
(\textbf{B}) Same as (A), but for loci pairs separated by less than the stated distance.}
\label{fig:hic-distance}
\end{figure*}

Previous work has used Hi-C data to investigate the short-range structure ($<500$ kb) of chromatin and understand how proteins 
like CTCF package DNA into loops called TADs~\cite{Tang2015}.
In \cref{fig:hic-distance}, we show that long-range contacts, rather than short-range contacts, arise as important for 
predicting cell type. This is significant because short-range contacts have been well-studied and are thought to be highly conserved 
between cell types and even species, whereas
less is known about the nature of long-range contacts.
To assess the contribution of long-range versus short-range contacts to predict cell type, we removed all contacts in a range either 
below (A) or above (B) the distance indicated by the legend. We observe that removal of all contacts $<500$ kb does not meaningfully 
impact the predictive accuracy, but removing contacts above this range causes accuracy to decrease. In addition, when keeping only 
the local contacts, the method is relatively poor at distinguishing cell types. Keeping contacts in the $500$--$1000$ kb range and 
the $2.5$--$10$ Mb range appears to enhance predictive accuracy.

To further substantiate whether Hi-C structures are different in different cell types, we employed a functional attribution method in 
which we removed the contacts of all loci associated with Variable-Diversity-Joining (VDJ) recombination 
($igH$, $igK$, and $igL$)~\cite{Fugmann2014} and re-ran our prediction model. Since VDJ is present in only the B cells in our 
dataset, the masking of this data should reduce the confidence of classifying B cells, which is exactly what happens. We 
calculated the number of B cell measurements whose prediction accuracy improves upon inclusion of VDJ loci and found that 
40 instances do under the correlation-based model with three eigenloci. We calculated a bootstrapped distribution by randomly 
selecting the same number of loci and examining how many total B cell measurements improved, and we found that the observed 
number is $>5$ SDs larger than the null expectation. This analysis demonstrates that (i) aspects of chromatin 
structure are cell-type and species specific, as these chromosomal regions are not conserved across organisms or human cell 
types and (ii) functional attribution is achievable by masking the loci and observing the change in probability.

\newpage
\FloatBarrier
%\clearpage
\section*{Supplementary Figures}

%FIG S1-CONFUSION FIG
\begin{figure*}[!htb]
\vspace{-1cm}
\centering
\includegraphics[width=.6\linewidth]{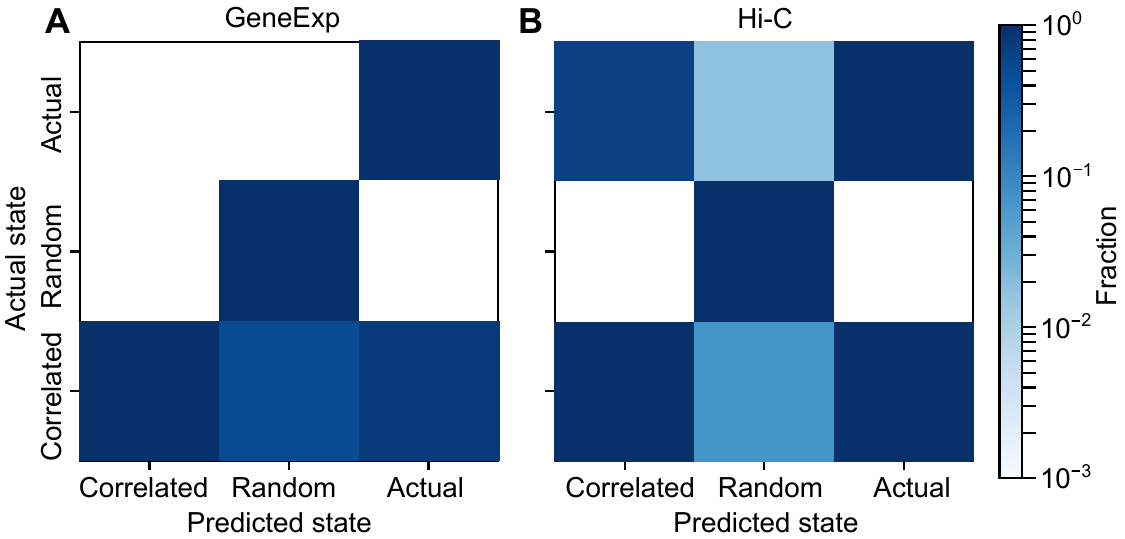}
\caption{\textbf{Confusion matrices for discerning actual and simulated data.}
(\textbf{A}) Distinguishability of actual data from the GeneExp dataset from uncorrelated simulated data (Uncorrelated), and correlated simulated 
data (Correlated), with accuracies color-coded as a fraction of the number of states predicted. 
In the confusion matrix, rows correspond to the actual method used to generate the data and 
columns map to the predicted method used to generate the data.
(\textbf{B}) Same as (A), but for the Hi-C dataset.
 In both datasets, actual data are confused with the 
simulated, correlated data much more frequently than they are with simulated, uncorrelated data. The 
misclassification rates are $>33\%$ for GeneExp and $>70\%$ for Hi-C.}
\label{fig:confusion}
\end{figure*}

%FIG S2-Comparison with PCA
\begin{figure*}[!htb]
 \centering
  \includegraphics[width=.6\linewidth]{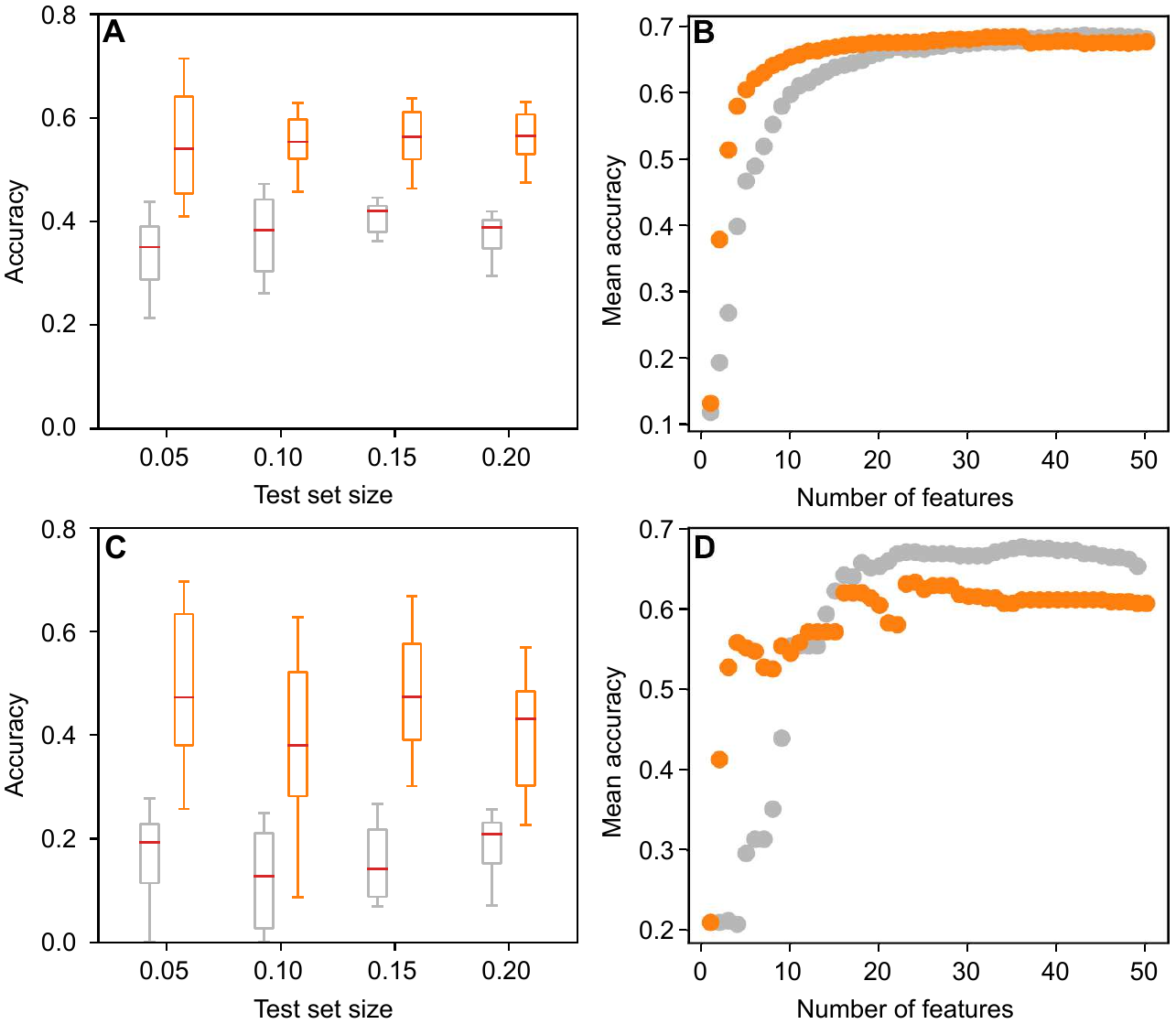}
 \caption{
 \textbf{Comparison of forward selection with PCA.}
 (\textbf{A}) Accuracy of PCA (grey) with forward selection (orange) as a function of test set size, expressed as a fraction of the total number of experiments in the GeneExp dataset.
 (\textbf{B}) Accuracy as a function of the number of features for the GeneExp dataset.
  (\textbf{C}) Same as (A), but for the Hi-C dataset.
 (\textbf{D}) Same as (B), but for the Hi-C dataset.
  Axes labels for (A)--(D) retain their meanings from \cref{fig:version_comparison}.
  Differences in all distributions in (A) and (C) are significant at the $p < 0.01$ level (Kolmogorov-Smirnov test). 
}
 \label{fig:pca_comp}
\end{figure*}
\newpage

%FIGS3-Hi-C pairwise distance distinguishability AND nonconvexity
\begin{figure*}
\centering
  \includegraphics[width=.67\linewidth]{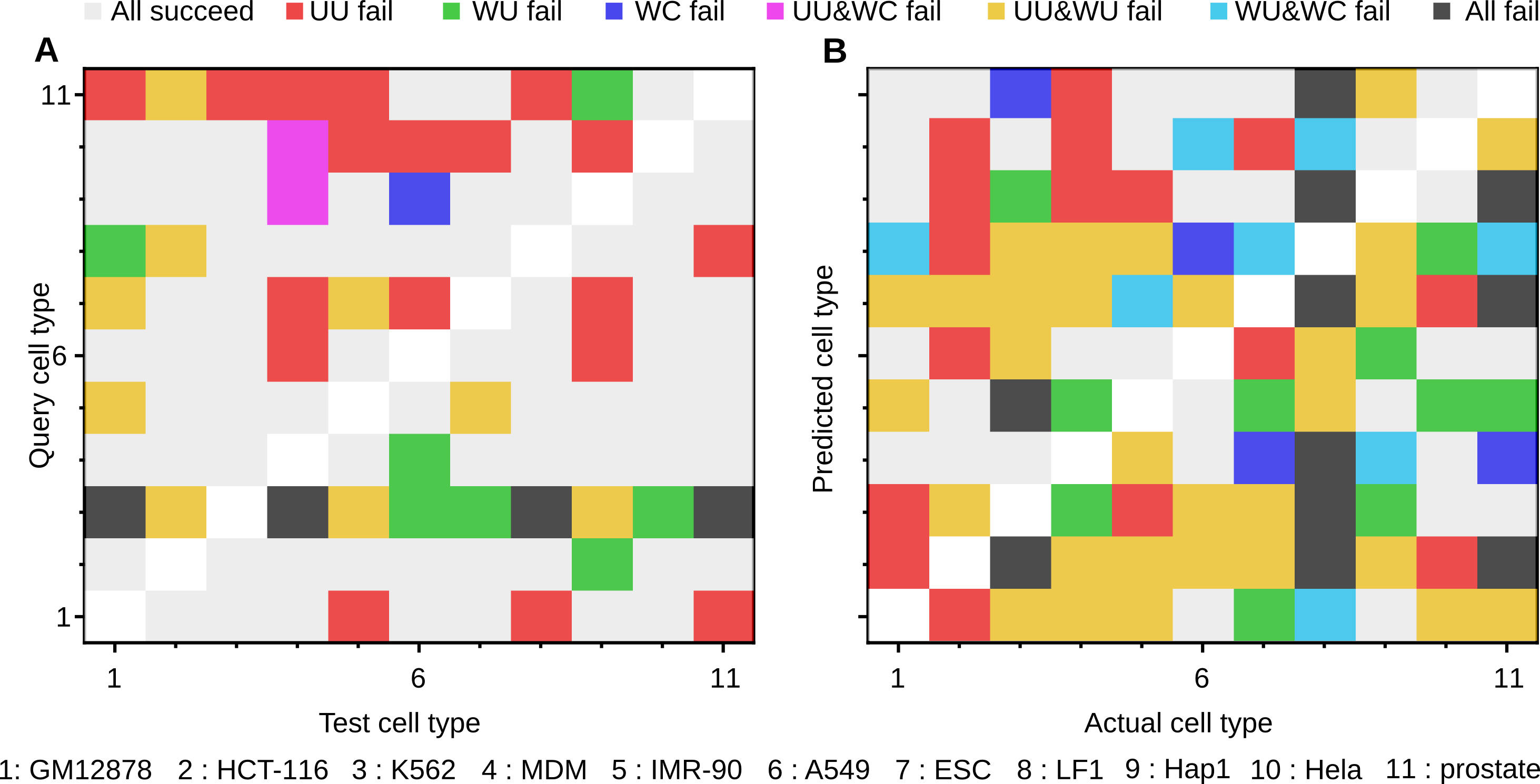}
 \caption{
 \textbf{Distinguishing cell types for the Hi-C dataset.} 
 (\textbf{A}) Cell type homogeneity, where axes and color code retain their meanings from \cref{fig:minsep-breakdown}. 
 (\textbf{B}) Nonconvex fraction for a sampling of chords between pairs of same cell type measurements with predicted cell types on the $y$ axis and actual cell types on the $x$ axis. Each square is colored if $> 0.1\%$ of chords of the actual cell type are classified as the predicted cell type using the versions of the method indicated in the legend.
 }
 \label{fig:hic-breakdown}
\end{figure*}

\newpage
%FIG S4-LOGO accuracy broken down by cell type
\begin{figure*}[htb]
  \centering
  \includegraphics[width=.8\linewidth]{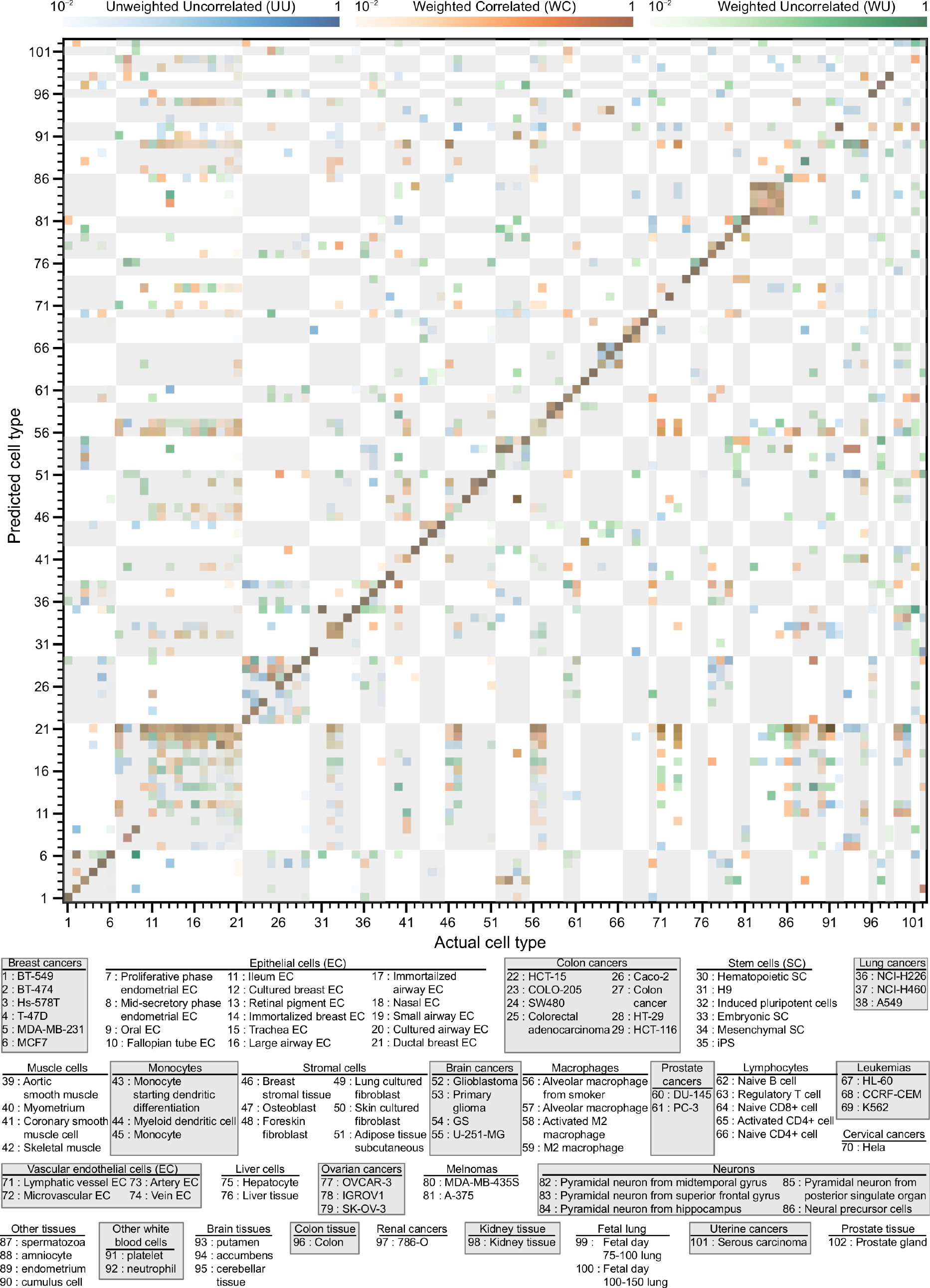}
 \caption{
 \textbf{KNN Classification accuracy by cell type for the GeneExp dataset under LOGO cross-validation.}
 Version abbreviations and colors bars are the defined in \cref{fig:knn_summary}A, and the grey and white checkered 
 background and tick label legend retain their meaning from \cref{fig:minsep-breakdown}. 
 %The number of experiments for each cell type are listed in Table~S2.
The accuracies averaged by cell type group correspond to those presented in \cref{fig:knn_summary}A. 
}
\label{fig:loo-detail}
\end{figure*}

\newpage
%FIG S5-Nonconvexity in the GeneExp dataset
\begin{figure*}[htb]
%    \vspace{-0.5cm}
    \centering
  \includegraphics[width=.8\linewidth]{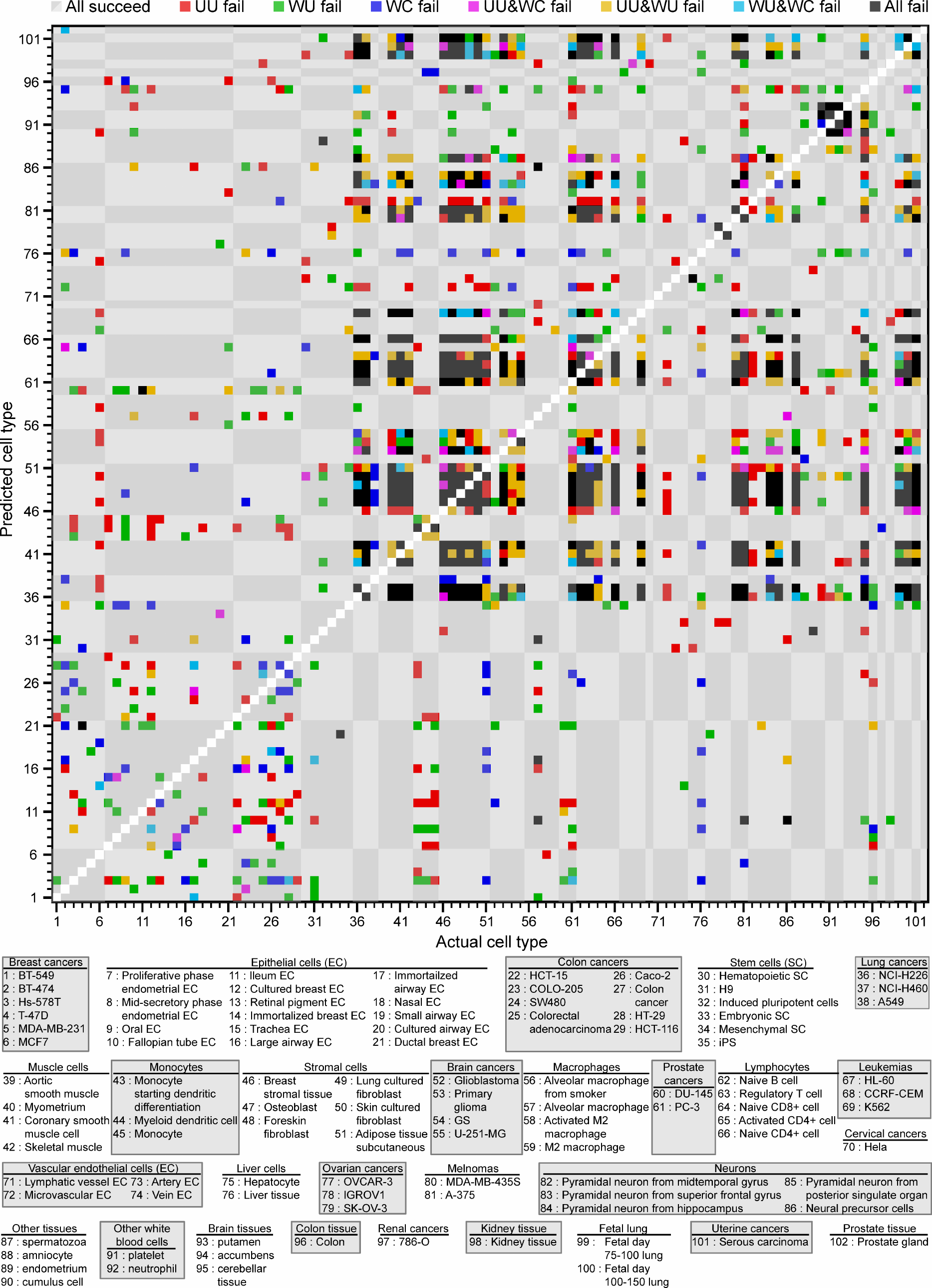}
 \caption{
\textbf{Fraction of nonconvex chords for each cell type.} 
Colors, background, abbreviations and 
tick label legend retain their values described in \cref{fig:minsep-breakdown}. The predicted cell 
type was distinguishable from the actual one if $ < 0.1\% $ of the chords from the actual 
cell type (column) were classified as the predicted cell type (row).}
\label{fig:ncv-GeneExp}
%\vspace{-0.5cm}
\end{figure*}

\newpage
%FIG S6-counts of cell type pairs that fail each criterion for each version for both datasets
\begin{figure*}[htb]
\centering
\includegraphics[width=.9\linewidth]{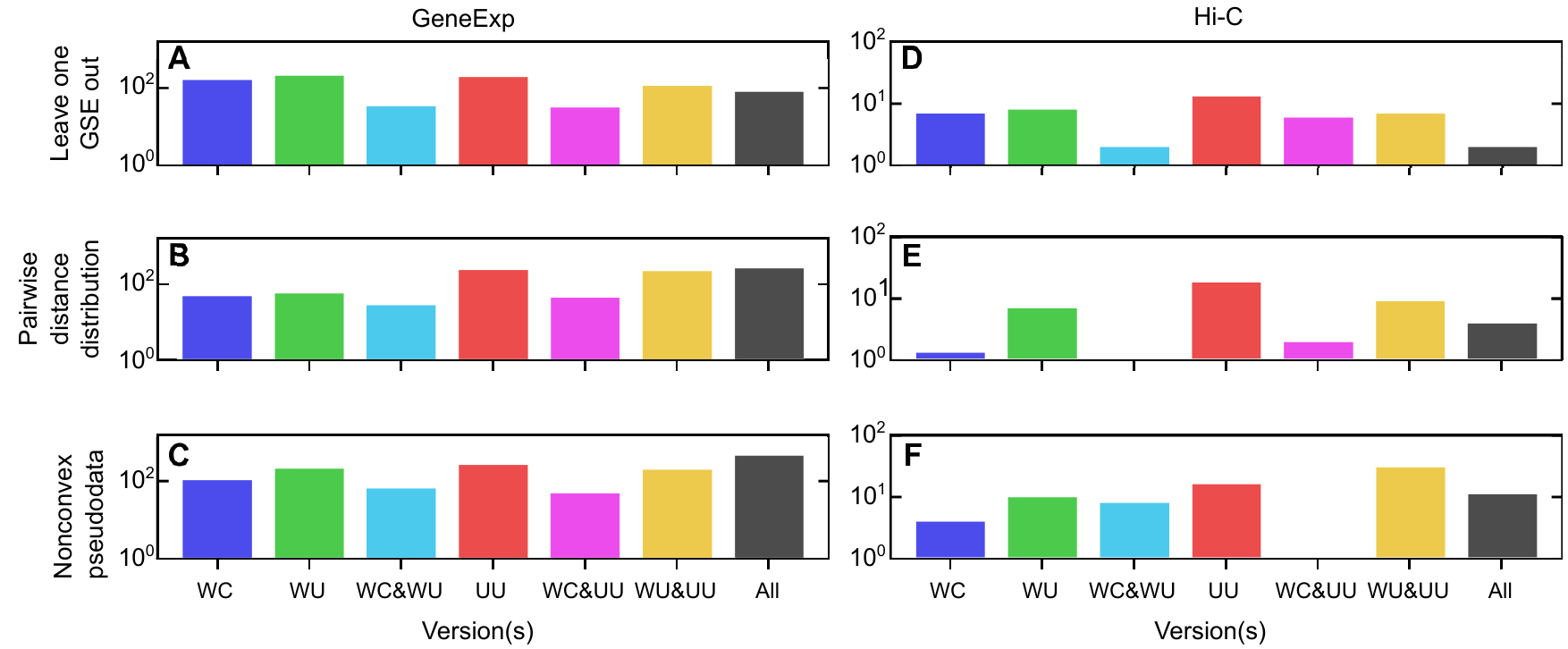}
 \captionof{figure}{
\textbf{Compilation of the number of squares of each color found in the preceding figures.} 
 (\textbf{A})~\cref{fig:loo-detail}. 
 (\textbf{B})~\cref{fig:minsep-breakdown}. 
 (\textbf{C})~\cref{fig:ncv-GeneExp}. 
 (\textbf{D})~\cref{fig:knn_summary}B. 
 (\textbf{E})~\cref{fig:hic-breakdown}A.
 (\textbf{F})~\cref{fig:hic-breakdown}B. 
 In (D), cell types are distinguishable (and therefore not counted) if $< 10\%$ of the experiments of a given cell type are classified as that specified by the $y$ axis. The fraction of cases in which all versions of our approach distinguish cell types are not shown to emphasize the differences between the versions.}
\label{fig:overlap-count}
\end{figure*}

\FloatBarrier
\clearpage
\section*{Supplementary Tables}

\begin{table}[h]
\caption{\label{tab:version-comparison}
Comparison between the three versions of the KNN technique on the three criteria measured by the percentage of cell type pairs distinguished for each dataset when limited to 4 (GeneExp) or 3 (Hi-C) features. (This corresponds to the ``Version Comparison'' sheet of ``Table S1.xlsx'' accompanying the published manuscript.) }
\begin{tabular}{ll | cccr}
Dataset & Criterion & UU$\,$(\%) & WU$\,$(\%) & WC$\,$(\%) & N \\ 
\hline
GeneExp  & \cref{eq:C1} & 92.8  &   94.6 &    96.4 & 10,302 \\
        &  \cref{eq:C2} & 51.8&   51.8 &   57.9 & 3,109 \\
        & \cref{eq:C3} &  72.5 &  72.4 &   77.5 & $1,020,000$\\
\hline
Hi-C & \cref{eq:C1} &   70.0  & 81.8 &  93.6 & 110\\
        & \cref{eq:C2} &   27.4 &  36.9 &  52.8 & 453 \\
        & \cref{eq:C3} &   68.6 &  66.1 &  89.5 & $110,000$ \\
\end{tabular}
\end{table}

\begin{table}[h]
\caption{\label{significance:tab}
Kolmogorov-Smirnov Test $p$-values for the box plots in \cref{fig:version_comparison}A, C. Column labels indicate the versions compared. (This corresponds to the ``KS test $p$-values'' sheet of ``Table S1.xlsx'' accompanying the published manuscript.)}
\begin{tabular}{l | ccc | ccc}
             &            &GeneExp&           &            &   Hi-C   & \\
Fraction&UU-WU&UU-WC&WC-WU&UU-WU&UU-WC&WC-WU\\
\hline
0.05      & 0.03    & 0.24    & 0.12  & 0.41 & 0.12 & 0.00 \\
0.10      & 0.24    & 0.24    & 0.01  & 0.24 & 0.03 & 0.00 \\
0.15      & 0.24    & 0.00    & 0.00  & 0.24 & 0.03 & 0.12 \\
0.20      & 0.65    & 0.03    & 0.00  & 0.03 & 0.00 & 0.00 \\
\end{tabular}
\end{table}

%
%\begin{table}[h]
%\centering
%\caption{\label{tab:version-comparison}
%\textbf{Version comparison results and KS test $p$-values.} Supplementary File: Supplementary Table S1.xlsx}
%\end{table} 
%
%\begin{table}[h]
%\caption{\label{supp_tab:ticklabels}
%\textbf{Cell type counts, tick labels for Figs. 2c, 3, 5, and S5--S6, and database accession numbers for the GeneExp and Hi-C datasets.} Supplementary File: Supplementary Table S2.xlsx}
%\end{table}


\begin{thebibliography}{10}
\providecommand{\bibnamefont}[1]{#1}
\providecommand{\bibfnamefont}[1]{#1}
\providecommand{\selectlanguage}[1]{\relax}

\bibitem{Cortini2016}
\bibfnamefont{R.}~\bibnamefont{Cortini}, \bibfnamefont{M.}~\bibnamefont{Barbi},
  \bibfnamefont{B.~R.} \bibnamefont{Car{\'{e}}},
  \bibfnamefont{C.}~\bibnamefont{Lavelle},
  \bibfnamefont{A.}~\bibnamefont{Lesne},
  \bibfnamefont{J.}~\bibnamefont{Mozziconacci}, and \bibfnamefont{J.-M.}
  \bibnamefont{Victor}, {The physics of epigenetics}. \emph{Rev. Mod. Phys.}
  \textbf{88}, 025002 (2016).

\bibitem{Shi2006}
\bibfnamefont{L.}~\bibnamefont{Shi}, \bibfnamefont{L.~H.} \bibnamefont{Reid},
  \bibfnamefont{W.~D.} \bibnamefont{Jones},
  \bibfnamefont{R.}~\bibnamefont{Shippy}, \bibfnamefont{J.~A.}
  \bibnamefont{Warrington}, \bibfnamefont{S.~C.} \bibnamefont{Baker},
  \bibfnamefont{P.~J.} \bibnamefont{Collins},
  \bibfnamefont{F.}~\bibnamefont{de~Longueville}, \bibfnamefont{E.~S.}
  \bibnamefont{Kawasaki}, \bibfnamefont{K.~Y.} \bibnamefont{Lee},
  \emph{et~al.}, {The MicroArray Quality Control (MAQC) project shows inter-
  and intraplatform reproducibility of gene expression measurements}. \emph{Nat.
  Biotechnol.} \textbf{24}, 1151--1161 (2006).

\bibitem{Imakaev2012}
\bibfnamefont{M.}~\bibnamefont{Imakaev},
  \bibfnamefont{G.}~\bibnamefont{Fudenberg}, \bibfnamefont{R.~P.}
  \bibnamefont{McCord}, \bibfnamefont{N.}~\bibnamefont{Naumova},
  \bibfnamefont{A.}~\bibnamefont{Goloborodko}, \bibfnamefont{B.~R.}
  \bibnamefont{Lajoie}, \bibfnamefont{J.}~\bibnamefont{Dekker}, and
  \bibfnamefont{L.~A.} \bibnamefont{Mirny}, {Iterative correction of Hi-C data
  reveals hallmarks of chromosome organization}. \emph{Nat. Methods} \textbf{9}, 999--1003
  (2012).

\bibitem{Ozsolak2010}
\bibfnamefont{F.}~\bibnamefont{Ozsolak} and \bibfnamefont{P.~M.}
  \bibnamefont{Milos}, {RNA sequencing: Advances, challenges and
  opportunities}. \emph{Nat. Rev. Genet.} \textbf{12}, 87--98 (2010).

\bibitem{Barrett2009}
\bibfnamefont{T.}~\bibnamefont{Barrett}, \bibfnamefont{D.~B.}
  \bibnamefont{Troup}, \bibfnamefont{S.~E.} \bibnamefont{Wilhite},
  \bibfnamefont{P.}~\bibnamefont{Ledoux},
  \bibfnamefont{D.}~\bibnamefont{Rudnev},
  \bibfnamefont{C.}~\bibnamefont{Evangelista}, \bibfnamefont{I.~F.}
  \bibnamefont{Kim}, \bibfnamefont{A.}~\bibnamefont{Soboleva},
  \bibfnamefont{M.}~\bibnamefont{Tomashevsky}, \bibfnamefont{K.~A.}
  \bibnamefont{Marshall}, \emph{et~al.}, {NCBI GEO: Archive for high-throughput
  functional genomic data}. \emph{Nucl. Acids Res.} \textbf{37}, D885--D890 (2009).

\bibitem{leinonen2010sequence}
\bibfnamefont{R.}~\bibnamefont{Leinonen},
  \bibfnamefont{H.}~\bibnamefont{Sugawara},
  \bibfnamefont{M.}~\bibnamefont{Shumway}, and \bibfnamefont{I.~N. S.~D.}
  \bibnamefont{Collaboration}, {The sequence read archive}. \emph{Nucl. Acids Res.}
  \textbf{39}, D19--D21 (2010).

\bibitem{Hopfield1982}
\bibfnamefont{J.~J.} \bibnamefont{Hopfield}, {Neural networks and physical
  systems with emergent collective computational abilities}. \emph{Proc. Natl. Acad.
  Sci. U.S.A.} \textbf{79}, 2554--2558 (1982).

\bibitem{Packard1980}
\bibfnamefont{N.~H.} \bibnamefont{Packard}, \bibfnamefont{J.~P.}
  \bibnamefont{Crutchfield}, \bibfnamefont{J.~D.} \bibnamefont{Farmer}, and
  \bibfnamefont{R.~S.} \bibnamefont{Shaw}, {Geometry from a time series}. \emph{Phys.
  Rev. Lett.} \textbf{45}, 712--716 (1980).

\bibitem{Lu2009}
\bibfnamefont{R.}~\bibnamefont{Lu}, \bibfnamefont{F.}~\bibnamefont{Markowetz},
  \bibfnamefont{R.~D.} \bibnamefont{Unwin}, \bibfnamefont{J.~T.}
  \bibnamefont{Leek}, \bibfnamefont{E.~M.} \bibnamefont{Airoldi},
  \bibfnamefont{B.~D.} \bibnamefont{MacArthur},
  \bibfnamefont{A.}~\bibnamefont{Lachmann},
  \bibfnamefont{R.}~\bibnamefont{Rozov},
  \bibfnamefont{A.}~\bibnamefont{Ma'ayan}, \bibfnamefont{L.~A.}
  \bibnamefont{Boyer}, \emph{et~al.}, {Systems-level dynamic analyses of fate
  change in murine embryonic stem cells}. \emph{Nature} \textbf{462}, 358--362 (2009).

\bibitem{Airoldi2016}
\bibfnamefont{E.~M.} \bibnamefont{Airoldi},
  \bibfnamefont{D.}~\bibnamefont{Miller},
  \bibfnamefont{R.}~\bibnamefont{Athanasiadou},
  \bibfnamefont{N.}~\bibnamefont{Brandt},
  \bibfnamefont{F.}~\bibnamefont{Abdul-Rahman},
  \bibfnamefont{B.}~\bibnamefont{Neymotin},
  \bibfnamefont{T.}~\bibnamefont{Hashimoto},
  \bibfnamefont{T.}~\bibnamefont{Bahmani}, and
  \bibfnamefont{D.}~\bibnamefont{Gresham}, {Steady-state and dynamic gene
  expression programs in \emph{Saccharomyces cerevisiae} in response to variation in
  environmental nitrogen}. \emph{Mol. Biol. Cell} \textbf{27}, 1383--1396 (2016).

\bibitem{Wytock2019}
\bibfnamefont{T.~P.} \bibnamefont{Wytock} and \bibfnamefont{A.~E.}
  \bibnamefont{Motter}, {Predicting growth rate from gene expression}. \emph{Proc.
  Natl. Acad. Sci. U.S.A.} \textbf{116}, 367--372 (2019).

\bibitem{Assaf2013}
\bibfnamefont{M.}~\bibnamefont{Assaf}, \bibfnamefont{E.}~\bibnamefont{Roberts},
  \bibfnamefont{Z.}~\bibnamefont{Luthey-Schulten}, and
  \bibfnamefont{N.}~\bibnamefont{Goldenfeld}, {Extrinsic noise driven phenotype
  switching in a self-regulating gene}. \emph{Phys. Rev. Lett.} \textbf{111}, 058102
  (2013).

\bibitem{Lu2014}
\bibfnamefont{M.}~\bibnamefont{Lu}, \bibfnamefont{J.}~\bibnamefont{Onuchic},
  and \bibfnamefont{E.}~\bibnamefont{Ben-Jacob}, {Construction of an effective
  landscape for multistate genetic switches}. \emph{Phys. Rev. Lett.} \textbf{113},
  078102 (2014).

\bibitem{Wang2012}
\bibfnamefont{R.-S.} \bibnamefont{Wang},
  \bibfnamefont{A.}~\bibnamefont{Saadatpour}, and
  \bibfnamefont{R.}~\bibnamefont{Albert}, {Boolean modeling in systems biology:
  An overview of methodology and applications}. \emph{Phys. Biol.} \textbf{9}, 055001
  (2012).

\bibitem{Saadatpour2016}
\bibfnamefont{A.}~\bibnamefont{Saadatpour} and
  \bibfnamefont{R.}~\bibnamefont{Albert}, {A comparative study of qualitative
  and quantitative dynamic models of biological regulatory networks}. \emph{EPJ
  Nonlinear Biomed. Phys.} \textbf{4}, 5 (2016).

\bibitem{Donner2010}
\bibfnamefont{R.~V.} \bibnamefont{Donner}, \bibfnamefont{Y.}~\bibnamefont{Zou},
  \bibfnamefont{J.~F.} \bibnamefont{Donges},
  \bibfnamefont{N.}~\bibnamefont{Marwan}, and
  \bibfnamefont{J.}~\bibnamefont{Kurths}, {Ambiguities in recurrence-based
  complex network representations of time series}. \emph{Phys. Rev. E} \textbf{81},
  015101 (2010).

\bibitem{Crutchfield2011}
\bibfnamefont{J.~P.} \bibnamefont{Crutchfield}, {Between order and chaos}. \emph{Nat.
  Phys.} \textbf{8}, 17--24 (2011).

\bibitem{Lang2014}
\bibfnamefont{A.~H.} \bibnamefont{Lang}, \bibfnamefont{H.}~\bibnamefont{Li},
  \bibfnamefont{J.~J.} \bibnamefont{Collins}, and
  \bibfnamefont{P.}~\bibnamefont{Mehta}, {Epigenetic landscapes explain
  partially reprogrammed cells and identify key reprogramming genes}. \emph{PLOS
  Comput. Biol.} \textbf{10}, e1003734 (2014).

\bibitem{Dettmer2016}
\bibfnamefont{S.~L.} \bibnamefont{Dettmer}, \bibfnamefont{H.~C.}
  \bibnamefont{Nguyen}, and \bibfnamefont{J.}~\bibnamefont{Berg}, {Network
  inference in the nonequilibrium steady state}. \emph{Phys. Rev. E} \textbf{94},
  052116 (2016).

\bibitem{Han2015}
\bibfnamefont{X.}~\bibnamefont{Han}, \bibfnamefont{Z.}~\bibnamefont{Shen},
  \bibfnamefont{W.-X.} \bibnamefont{Wang}, and
  \bibfnamefont{Z.}~\bibnamefont{Di}, {Robust Reconstruction of Complex
  Networks from Sparse Data}. \emph{Phys. Rev. Lett.} \textbf{114}, 028701 (2015).

\bibitem{Rondelez2012}
\bibfnamefont{Y.}~\bibnamefont{Rondelez}, {Competition for catalytic resources
  alters biological network dynamics}. \emph{Phys. Rev. Lett.} \textbf{108}, 018102
  (2012).

\bibitem{Braun2011}
\bibfnamefont{R.}~\bibnamefont{Braun}, \bibfnamefont{G.}~\bibnamefont{Leibon},
  \bibfnamefont{S.}~\bibnamefont{Pauls}, and
  \bibfnamefont{D.}~\bibnamefont{Rockmore}, {Partition decoupling for
  multi-gene analysis of gene expression profiling data}. \emph{BMC Bioinformatics}
  \textbf{12}, 497 (2011).

\bibitem{Kiselev2017}
\bibfnamefont{V.~Y.} \bibnamefont{Kiselev},
  \bibfnamefont{K.}~\bibnamefont{Kirschner}, \bibfnamefont{M.~T.}
  \bibnamefont{Schaub}, \bibfnamefont{T.}~\bibnamefont{Andrews},
  \bibfnamefont{A.}~\bibnamefont{Yiu}, \bibfnamefont{T.}~\bibnamefont{Chandra},
  \bibfnamefont{K.~N.} \bibnamefont{Natarajan},
  \bibfnamefont{W.}~\bibnamefont{Reik},
  \bibfnamefont{M.}~\bibnamefont{Barahona}, \bibfnamefont{A.~R.}
  \bibnamefont{Green}, \emph{et~al.}, {SC3: Consensus clustering of single-cell
  RNA-seq data}. \emph{Nat. Methods} \textbf{14}, 483--486 (2017).

\bibitem{Crow2018}
\bibfnamefont{M.}~\bibnamefont{Crow}, \bibfnamefont{A.}~\bibnamefont{Paul},
  \bibfnamefont{S.}~\bibnamefont{Ballouz}, \bibfnamefont{Z.~J.}
  \bibnamefont{Huang}, and \bibfnamefont{J.}~\bibnamefont{Gillis},
  {Characterizing the replicability of cell types defined by single cell
  RNA-sequencing data using MetaNeighbor}. \emph{Nat. Commun.} \textbf{9}, 884 (2018).

\bibitem{Kim2009}
\bibfnamefont{J.~B.} \bibnamefont{Kim}, \bibfnamefont{B.}~\bibnamefont{Greber},
  \bibfnamefont{M.~J.} \bibnamefont{Ara{\'{u}}zo-Bravo},
  \bibfnamefont{J.}~\bibnamefont{Meyer}, \bibfnamefont{K.~I.}
  \bibnamefont{Park}, \bibfnamefont{H.}~\bibnamefont{Zaehres}, and
  \bibfnamefont{H.~R.} \bibnamefont{Sch{\"{o}}ler}, {Direct reprogramming of
  human neural stem cells by OCT4}. \emph{Nature} \textbf{461}, 649--653 (2009).

\bibitem{Schwanhausser2011}
\bibfnamefont{B.}~\bibnamefont{Schwanh{\"{a}}usser},
  \bibfnamefont{D.}~\bibnamefont{Busse}, \bibfnamefont{N.}~\bibnamefont{Li},
  \bibfnamefont{G.}~\bibnamefont{Dittmar},
  \bibfnamefont{J.}~\bibnamefont{Schuchhardt},
  \bibfnamefont{J.}~\bibnamefont{Wolf}, \bibfnamefont{W.}~\bibnamefont{Chen},
  and \bibfnamefont{M.}~\bibnamefont{Selbach}, {Global quantification of
  mammalian gene expression control}. \emph{Nature} \textbf{473}, 337--342 (2011).

\bibitem{Tang2015}
\bibfnamefont{Z.}~\bibnamefont{Tang}, \bibfnamefont{O.~J.} \bibnamefont{Luo},
  \bibfnamefont{X.}~\bibnamefont{Li}, \bibfnamefont{M.}~\bibnamefont{Zheng},
  \bibfnamefont{J.~J.} \bibnamefont{Zhu},
  \bibfnamefont{P.}~\bibnamefont{Szalaj},
  \bibfnamefont{P.}~\bibnamefont{Trzaskoma},
  \bibfnamefont{A.}~\bibnamefont{Magalska},
  \bibfnamefont{J.}~\bibnamefont{Wlodarczyk},
  \bibfnamefont{B.}~\bibnamefont{Ruszczycki}, \emph{et~al.}, {CTCF-mediated
  human 3D genome architecture reveals chromatin topology for transcription}.
  \emph{Cell} \textbf{163}, 1611--1627 (2015).

\bibitem{ernst2011mapping}
\bibfnamefont{J.}~\bibnamefont{Ernst},
  \bibfnamefont{P.}~\bibnamefont{Kheradpour}, \bibfnamefont{T.~S.}
  \bibnamefont{Mikkelsen}, \bibfnamefont{N.}~\bibnamefont{Shoresh},
  \bibfnamefont{L.~D.} \bibnamefont{Ward}, \bibfnamefont{C.~B.}
  \bibnamefont{Epstein}, \bibfnamefont{X.}~\bibnamefont{Zhang},
  \bibfnamefont{L.}~\bibnamefont{Wang}, \bibfnamefont{R.}~\bibnamefont{Issner},
  \bibfnamefont{M.}~\bibnamefont{Coyne}, \emph{et~al.}, {Mapping and analysis
  of chromatin state dynamics in nine human cell types}. \emph{Nature} \textbf{473},
  43--49 (2011).

\bibitem{Marco2017}
\bibfnamefont{E.}~\bibnamefont{Marco},
  \bibfnamefont{W.}~\bibnamefont{Meuleman},
  \bibfnamefont{J.}~\bibnamefont{Huang}, \bibfnamefont{K.}~\bibnamefont{Glass},
  \bibfnamefont{L.}~\bibnamefont{Pinello},
  \bibfnamefont{J.}~\bibnamefont{Wang}, \bibfnamefont{M.}~\bibnamefont{Kellis},
  and \bibfnamefont{G.-C.} \bibnamefont{Yuan}, {Multi-scale chromatin state
  annotation using a hierarchical hidden Markov model}. \emph{Nat. Commun.}
  \textbf{8}, 15011 (2017).

\bibitem{Haury2011}
\bibfnamefont{A.-C.} \bibnamefont{Haury},
  \bibfnamefont{P.}~\bibnamefont{Gestraud}, and \bibfnamefont{J.-P.}
  \bibnamefont{Vert}, {The influence of feature selection methods on accuracy,
  stability and interpretability of molecular signatures}. \emph{PLOS ONE} \textbf{6},
  e28210 (2011).

\bibitem{Transtrum2011}
\bibfnamefont{M.~K.} \bibnamefont{Transtrum}, \bibfnamefont{B.~B.}
  \bibnamefont{Machta}, and \bibfnamefont{J.~P.} \bibnamefont{Sethna},
  {Geometry of nonlinear least squares with applications to sloppy models and
  optimization}. \emph{Phys. Rev. E} \textbf{83}, 036701 (2011).

\bibitem{yang2012network}
\bibfnamefont{Y.}~\bibnamefont{Yang}, \bibfnamefont{J.}~\bibnamefont{Wang}, and
  \bibfnamefont{A.~E.} \bibnamefont{Motter}, {Network observability
  transitions}. \emph{Phys. Rev. Lett.} \textbf{109}, 258701 (2012).

\bibitem{Bell1997}
\bibfnamefont{G.}~\bibnamefont{Bell} and \bibfnamefont{A.~O.}
  \bibnamefont{Mooers}, Size and complexity among multicellular organisms.
  \emph{Biol. J. Linn. Soc.} \textbf{60}, 345--363 (1997).

\bibitem{Bonner2004}
\bibfnamefont{J.~T.} \bibnamefont{Bonner}, {Perspective: the size-complexity
  rule}. \emph{Evolution} \textbf{58}, 1883--1890 (2004).

\bibitem{Huang2005}
\bibfnamefont{S.}~\bibnamefont{Huang}, \bibfnamefont{G.}~\bibnamefont{Eichler},
  \bibfnamefont{Y.}~\bibnamefont{Bar-Yam}, and \bibfnamefont{D.~E.}
  \bibnamefont{Ingber}, {Cell fates as high-dimensional attractor states of a
  complex gene regulatory network}. \emph{Phys. Rev. Lett.} \textbf{94}, 128701
  (2005).

\bibitem{Sarropoulos2019}
\bibfnamefont{I.}~\bibnamefont{Sarropoulos},
  \bibfnamefont{R.}~\bibnamefont{Marin},
  \bibfnamefont{M.}~\bibnamefont{Cardoso-Moreira}, and
  \bibfnamefont{H.}~\bibnamefont{Kaessmann}, {Developmental dynamics of lncRNAs
  across mammalian organs and species}. \emph{Nature} \textbf{571}, 510--514 (2019).

\bibitem{Horvat2015}
\bibfnamefont{S.}~\bibnamefont{Horv{\'{a}}t},
  \bibfnamefont{{\'{E}}.}~\bibnamefont{Czabarka}, and
  \bibfnamefont{Z.}~\bibnamefont{Toroczkai}, {Reducing degeneracy in maximum
  entropy models of networks}. \emph{Phys. Rev. Lett.} \textbf{114}, 158701 (2015).

\bibitem{Maaten2008}
\bibfnamefont{L.~v.~d.} \bibnamefont{Maaten} and
  \bibfnamefont{G.}~\bibnamefont{Hinton}, {Visualizing data using t-SNE}. 
  \emph{J. Mach. Learn. Res.} \textbf{9}, 2579--2605 (2008).

\bibitem{Sonawane2019}
\bibfnamefont{A.~R.} \bibnamefont{Sonawane}, \bibfnamefont{S.~T.}
  \bibnamefont{Weiss}, \bibfnamefont{K.}~\bibnamefont{Glass}, and
  \bibfnamefont{A.}~\bibnamefont{Sharma}, {Network medicine in the age of
  biomedical big data}. \emph{Front. Genet.} \textbf{10}, 294 (2019).

\bibitem{Dai2005}
\bibfnamefont{M.}~\bibnamefont{Dai}, \bibfnamefont{P.}~\bibnamefont{Wang},
  \bibfnamefont{A.~D.} \bibnamefont{Boyd},
  \bibfnamefont{G.}~\bibnamefont{Kostov},
  \bibfnamefont{B.}~\bibnamefont{Athey}, \bibfnamefont{E.~G.}
  \bibnamefont{Jones}, \bibfnamefont{W.~E.} \bibnamefont{Bunney},
  \bibfnamefont{R.~M.} \bibnamefont{Myers}, \bibfnamefont{T.~P.}
  \bibnamefont{Speed}, \bibfnamefont{H.}~\bibnamefont{Akil}, \emph{et~al.},
  {Evolving gene/transcript definitions significantly alter the interpretation
  of GeneChip data}. \emph{Nucl. Acids Res.} \textbf{33}, e175 (2005).

\bibitem{Irizarry2003}
\bibfnamefont{R.~A.} \bibnamefont{Irizarry},
  \bibfnamefont{B.}~\bibnamefont{Hobbs},
  \bibfnamefont{F.}~\bibnamefont{Collin}, \bibfnamefont{Y.~D.}
  \bibnamefont{Beazer‐Barclay}, \bibfnamefont{K.~J.}
  \bibnamefont{Antonellis}, \bibfnamefont{U.}~\bibnamefont{Scherf}, and
  \bibfnamefont{T.~P.} \bibnamefont{Speed}, {Exploration, normalization, and
  summaries of high density oligonucleotide array probe level data}.
\emph{Biostatistics} \textbf{4}, 249--264 (2003).

\bibitem{Johnson2007}
\bibfnamefont{W.~E.} \bibnamefont{Johnson}, \bibfnamefont{C.}~\bibnamefont{Li},
  and \bibfnamefont{A.}~\bibnamefont{Rabinovic}, {Adjusting batch effects in
  microarray expression data using empirical Bayes methods}. \emph{Biostatistics}
  \textbf{8}, 118--127 (2007).

\bibitem{Fugmann2014}
\bibfnamefont{S.~D.} \bibnamefont{Fugmann}, {Form follows function --- The
  three-dimensional structure of antigen receptor gene loci}. \emph{Curr. Opin.
  Immunol.} \textbf{27}, 33--37  (2014).

\end{thebibliography}
\end{document}